\documentclass[10pt, journal, compsoc, onecolumn]{IEEEtran}

\usepackage{textcomp}
\usepackage{hyperref}
\hypersetup{anchorcolor=black,citecolor=black}
\usepackage{tikz}
\usepackage{amsthm}
\usepackage{amsmath}
\usepackage{mathtools}
\usepackage{adjustbox}
\usepackage[linesnumbered,ruled,vlined]{algorithm2e}
\usepackage{graphicx}
\newcommand{\mygrid}{\tikz{ \draw [step=5pt, thick]  (6.1,4) grid  (9,6);}}
\usepackage{url}
\usetikzlibrary{arrows,positioning, calc}
\tikzstyle{vertex}=[draw,fill=black!15,circle,minimum size=18pt,inner sep=0pt]
\usepackage{amsmath}
\usepackage{multicol}
\usepackage{multirow}
\usepackage{pgfplots}
\pgfplotsset{compat=newest}
\usepackage[utf8]{inputenc}
\usepackage[FIGBOTCAP]{subfigure}
\usepackage{listings, lstautogobble}
\usepackage{diagbox}
\usepackage{float}
\usepackage{hhline}
\usepackage{mathtools}
\usepackage{enumitem}
\usetikzlibrary{positioning,shapes}
\makeatletter
\def\url@leostyle{%
  \@ifundefined{selectfont}{\def\UrlFont{\sf}}{\def\UrlFont{\small\ttfamily}}}
\makeatother
\pgfplotsset{every axis/.append style={
                    xlabel={$x$},          
                    ylabel={$y$},          
                    label style={font=\Large},
                    tick label style={font=\Large},
                    legend style ={font=\boldmath}
                    }}

 \AtBeginDocument{
   \addtolength{\abovedisplayskip}{-12pt}
   \addtolength{\abovedisplayshortskip}{-12pt}
  \addtolength{\belowdisplayskip}{-10pt}
   \addtolength{\belowdisplayshortskip}{-10pt}
   \addtolength{\textfloatsep}{-7pt}
   \addtolength{\dblfloatsep}{-7pt}
   \addtolength{\abovecaptionskip}{-3pt}
   \addtolength{\belowcaptionskip}{-3pt}
 }
\pgfplotsset{compat=1.14}

\begin{document}

\bstctlcite{IEEEexample:BSTcontrol}

\title{Tracking Normalized Network Traffic Entropy \\ to Detect DDoS Attacks in P4}
	
\author{Damu Ding\href{https://orcid.org/0000-0001-9692-7756}{\protect\includegraphics[scale=0.1]{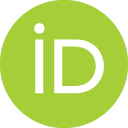}},~ \IEEEmembership{Student Member, ~IEEE},
        Marco Savi\href{https://orcid.org/0000-0002-8193-0597}{\protect\includegraphics[scale=0.1]{orcid/icon.png}},
        and Domenico Siracusa\href{https://orcid.org/0000-0002-5640-6507}{\protect\includegraphics[scale=0.1]{orcid/icon.png}} 
\thanks{Damu Ding was with Fondazione Bruno Kessler, Trento, Italy and University of Bologna, Bologna, Italy. He is now with University of Oxford. E-mail: damu.ding@eng.ox.ac.uk.  Marco Savi is with University of Milano-Bicocca, Milano, Italy. E-mail: marco.savi@unimib.it. Domenico Siracusa is with Fondazione Bruno Kessler,  Trento, Italy. E-mail:  dsiracusa@fbk.eu.A preliminary version of this paper appeared in \protect\cite{ding2019estimating}, presented at IEEE/IFIP NOMS in 2020. The research leading to these results has received funding from the EC within the H2020 Research and Innovation program, Grant Agreement No. 856726 (GN4-3 project).}
}
\IEEEtitleabstractindextext{%
\begin{abstract}
Distributed Denial-of-Service (DDoS) attacks represent a persistent threat to modern telecommunications networks: detecting and counteracting them is still a crucial unresolved challenge for network operators. DDoS attack detection is usually carried out in one or more central nodes that collect significant amounts of monitoring data from networking devices, potentially creating issues related to network overload or delay in detection. The dawn of programmable data planes in Software-Defined Networks can help mitigate this issue, opening the door to the detection of DDoS attacks directly in the data plane of the switches. However, the most widely-adopted data plane programming language, namely P4, lacks supporting many arithmetic operations, therefore, some of the advanced network monitoring functionalities needed for DDoS detection cannot be straightforwardly implemented in P4. This work overcomes such a limitation and presents two novel strategies for flow cardinality and for normalized network traffic entropy estimation that only use P4-supported operations and guarantee a low relative error. Additionally, based on these contributions, we propose a DDoS detection strategy relying on variations of the normalized network traffic entropy. Results show that it has comparable or higher detection accuracy than state-of-the-art solutions, yet being simpler and entirely executed in the data plane. 

\end{abstract}

\begin{IEEEkeywords}
Network monitoring, Programmable data planes, P4, Normalized network traffic entropy, DDoS detection
\end{IEEEkeywords}}
\maketitle



\section{Introduction}

\IEEEPARstart{D}{istributed} Denial-of-Service (DDoS) attacks are becoming one of the most significant threats for network operators and their customers as such attacks, carried out by many different compromised hosts, are able to flood a victim with a huge load of superfluous traffic and exhaust its network and computational resources, causing service disruptions. In this context, detecting DDoS attacks in a smooth yet effective way plays a key role in today's network security. Periodical monitoring of specific network metrics has been widely adopted as a strategy to detect DDoS attacks. For instance, network traffic entropy is a statistical measure to describe the flow distribution, and the entropy of distinct destination IPs observed in the network significantly decreases during a DDoS attack \cite{giotis2014combining}\cite{kalkan2018jess}\cite{wang2015entropy}. Moreover, a significant increase in the number of source IPs contacting a specific destination IP  \cite{yu2013software}\cite{liu2016one}\cite{huang2017sketchvisor} may also indicate that a DDoS attack is taking place.

From a technological perspective, both in SNMP-based \cite{van2014opennetmon} networks and in more recent Openflow-based \cite{mckeown08} Software-Defined Networks (SDNs), monitoring data collection and, consequently, DDoS detection are carried out by a logically centralized component (generally known as \emph{monitoring collector} or, more widely, \emph{controller}): this requires the transmission, storage,  and processing of a large amount of information related to the network state from network devices to this component \cite{sivaraman2017heavy}. Such an approach comes with two well-known drawbacks \cite{ding2019incremental}: \emph{(i)} a significant communication overhead is generated between data and centralized monitoring/control planes and \emph{(ii)} significant processing capabilities are needed by the collector, with the risk of affecting the performance of monitoring and network operations if involved parties are not well-dimensioned.

The recent advent of so-called (data-plane) \emph{programmable switches} allows network operators to partially overcome such drawbacks. In fact, programmable switches can, if appropriately programmed, execute part of the network monitoring/security operations directly in their data plane pipeline and  deliver to the centralized monitoring/control plane information that is partially or fully processed. 
However, data-plane programming comes with some inherent limitations: the most well-established and widely-adopted data-plane programming language, called P4 \cite{bosshart2014p4}, does not natively support basic yet relevant arithmetic operations such as division, logarithm and exponential function calculation, 
as well as any operation on floating numbers or \emph{for} loops. 
Unfortunately, all these operations are needed to effectively implement an \emph{entropy-based DDoS detection} strategy that \emph{(i)} is able to evaluate abnormal variations on the entropy over time and \emph{(ii)} can be fully executed in the programmable data plane, that is, it operates within the data plane pipeline and forwards alarms to the  monitoring collector to notify about potential DDoS attacks. 

However, spotting  variations of entropy over time, as done in previous works, may not be the most effective way to detect DDoS attacks. In fact, the number of distinct flows in the network (i.e., \emph{flow cardinality}) changes dynamically, affecting, in turn, the value of traffic entropy. A more suitable metric is therefore the \emph{normalized entropy}, which is normalized against flow cardinality and is  more robust to legitimate changes on the number of distinct flows. 


The goal of this paper is thus to propose novel strategies to estimate network traffic statistics such as normalized entropy and flow cardinality directly in P4 programmable switches, with the final goal of using them as building blocks to accurately and timely detect DDoS attacks.
To this aim, based on  P4-based solutions for the estimation of logarithm (P4Log) and exponential function (P4Exp) that we proposed in a preliminary version of this work \cite{ding2019estimating}, we here propose \emph{P4LogLog}, a novel memory-efficient strategy that takes inspiration from LogLog algorithm \cite{durand2003loglog} for the estimation of flow cardinality in P4.
We then present \emph{P4NEntropy}, our strategy for normalized Shannon entropy \cite{shannon2001mathematical} estimation in P4, and \emph{P4DDoS}, our approach for DDoS detection based on P4NEntropy. 
Even though we designed and implemented P4LogLog and P4NEntropy in support to P4DDoS, they can be seen as two stand-alone strategies paving the way towards the development of new monitoring capabilities in programmable data planes. The prototypes of P4LogLog, P4NEntropy and P4DDoS have been implemented with the P4 behavioral model \cite{p4simulator} and proved to be fully executable in a P4 emulated environment.

We then evaluate P4LogLog, P4NEntropy and P4DDoS by means of simulations to show their effectiveness and their sensitivity to different tuning parameters, with three critical improvements (to the best of our knowledge) with respect to the literature: 
\begin{itemize} [leftmargin=*]
\item P4LogLog can guarantee better accuracy than a widely-adopted state-of-the-art flow cardinality estimator \cite{yang2018elastic} while ensuring small memory usage. 
\item P4NEntropy ensures a comparable relative error in entropy estimation to a P4-based state-of-the-art solution \cite{lapolli2019offloading}, but it avoids the usage of pre-computed values stored in the Ternary Content-Addressable Memory (TCAM) and adopts a time-interval-based window instead of a packet-based one, which eases the switches' synchronization if additional network-wide operations should be executed.  
\item P4DDoS ensures slightly better performance than an existing P4-enabled entropy-based DDoS detection solution \cite{lapolli2019offloading}. In case of some stealthy DDoS attacks, such as an internal botnet DDoS attack or a DDoS attack with spoofed source IPs, our P4DDoS outperforms the state-of-the-art solution in terms of detection accuracy. Moreover, our P4DDoS does not need any interaction with the control plane in executing the foreseen operations, whereas \cite{lapolli2019offloading} requires that a controller properly populates the TCAM of the switch with some pre-computed values. This is why we claim that \emph{our strategy works entirely in the data plane}. 
\end{itemize}


The remainder of the paper is organized as follows. In Section \ref{sec:background} we report background notions. Section \ref{sec:hint} motivates why we choose P4Log and P4Exp (see \cite{ding2019estimating}) as building blocks for P4LogLog and P4NEntropy. Section \ref{sec:nentropy} describes P4LogLog and P4NEntropy, while Section \ref{sec:ddos} describes P4DDoS. 
Sections \ref{sec:evaluation1} and \ref{sec:evaluation2} present evaluation results and comparisons with existing solutions. In Section \ref{sec:related}  we recall the related work. Finally, Section \ref{sec:conclusion} concludes the paper and discusses the future work.

\section{Background} \label{sec:background}
In this section we recall background concepts needed to understand the strategies proposed in the following sections.
\subsection{Normalized network traffic entropy} \label{sec:entropy}
Network traffic entropy \cite{lall2006data} gives an indication on traffic distribution across the network. Each network switch can evaluate the traffic entropy related to the network flows that cross it in a given time interval $T_{int}$. Relying on the definition of \emph{Shannon entropy} \cite{shannon2001mathematical}, network traffic entropy can be defined as $H = -\sum_{i=1}^{n}{\frac{f_{i}}{|S|_{tot}} \log_{d}{\frac{f_{i}}{|S|_{tot}}}}$, where $f_{i}$ is the packet count of the incoming flow with \emph{flow key} $i$ (e.g. 5 tuple, source IP-destination IP pair, etc.),  $|S|_{tot}$ is the total number of processed packets by the switch during $T_{int}$, $n$ is the overall number of distinct flows and $d$ is the base of logarithm. Traffic entropy is $H=0$ when in $T_{int}$ all packets $|S|_{tot}$ belong to the same flow $i$, while it assumes its maximum value $H=\log_d n$ when packets are uniformly distributed among the $n$ flows. The \emph{normalized entropy} is defined as $H_{norm} = \frac{H}{\log_d(n)}$ ($0\le H_{norm}\le 1$). 

\subsection{Hamming weight computation} \label{hamming}
Hamming weight represents the number of non-zero values in a string. In a binary string, the Hamming weight indicates the overall number of ones. For example, given the  binary string 01101, the Hamming weight is 3. It can be computed by means of different algorithms: as part of P4LogLog, in this paper we adopt the \emph{Counting 1-Bits} algorithm presented in \cite{warren2013hacker}, as it only relies on bitwise operations that are completely supported by the P4 language \cite{p4oper}.

\subsection{Sketch-based estimation of flow packet count} \label{sec:background_sketch}
Estimating the number of packets for a specific flow crossing a programmable switch ($f_i$) is fundamental for network traffic entropy computation. 
Such an estimation can be performed by means of \emph{sketches} \cite{huang2017sketchvisor}, which are probabilistic data structures associated to a set of pairwise-independent hash functions. The \emph{size} of each sketch data structure depends on the number of associated hash functions $N_{h}$ and on the output size of each function $N_{s}$, and is $N_{h}\times N_{s}$. 
\emph{Update} and \emph{Query} operations are used to store and retrieve information from the sketch: Update operation is responsible for updating the sketch to keep track of flow packet counts, while Query operation retrieves the estimated number of packets for a specific flow. Two well-known algorithms to Update and Query sketches are \emph{Count-min Sketch} \cite{cormode2009count} and \emph{Count Sketch} \cite{charikar2002finding}. A detailed theoretical analysis on the accuracy/memory occupation trade-off for these sketching algorithms is reported in \cite{cormode2009count}\cite{charikar2002finding}. From a high-level perspective, as any of $N_h$ and $N_s$ increase, memory consumption is larger but estimation is more accurate. Count Sketch leads to a better accuracy/memory consumption trade-off than Count-min Sketch, but its update time is twice slower \cite{cormode2011sketch}.

\subsection{LogLog algorithm for flow cardinality estimation} \label{background:distinct}
 \emph{LogLog} \cite{durand2003loglog} is a sketch-based algorithm that can be adopted to estimate the number of distinct flows crossing a switch.
 In brief, it works as follows. Given an incoming packet with \emph{flow key} $i$, LogLog applies to $i$ a hash function with output size $os$: the resulted $os$-bit binary string $s$ is denoted by $s = \{s_{os-1}s_{os-2}\cdots s_{0}\}$. LogLog then updates an $m$-sized LogLog register $Reg$. Let $bucket$ be the rightmost $k$ bits of $s$ (with $k = \log_{2}{m}$) and $x$ the remaining bits, i.e.,  $bucket = \{s_{k-1} \cdots s_{0}\}$ and $x=\{s_{os-1}\cdots s_{k}\}$. $Reg$ is updated following this rule: $Reg[bucket] = max(Reg[bucket], value)$, where $value$ is the index of the rightmost 1 of $x$ plus one. 
 $Reg$ can then be queried to estimate the flow cardinality $\hat{n}$, which is computed as $\hat{n}$ = $\alpha_{m}m 2^{\frac{1}{m}\sum_{bucket=0}^{m-1}{Reg[bucket]}}$, where $\alpha_{m}$ is a bias correction parameter. 
  An interesting property of LogLog is that multiple LogLog sketches can be merged to a single sketch, which can be used to count the flow cardinality of the union of many packet streams.

\section{Comparison of Log and Exp estimation strategies in programmable data planes} \label{sec:hint}

\begin{table}
 \caption{P4 programs properties}
\label{tuning}
 \centering
\resizebox{0.49\textwidth}{!}{
\begin{tabular}{|c|c|c|c|c|}
\hline
\textbf{Algorithm} &\textbf{Parameter} \cite{ding2019estimating} & \textbf{Value} \cite{ding2019estimating} & \textbf{Instructions} & \textbf{M+A entries}\\
\hline
\multirow{2}{*}{P4Log} & $N_{digits}$ & 3& \multirow{2}{*}{47} & \multirow{2}{*}{1}\\
\cline{2-3}
~  & $N_{bits}$ & 4 & ~ & ~\\
\hline
\multirow{1}{*}{P4Exp}  & $N_{terms}$  & 7  & 64 & 1\\ 
\hline
\multirow{1}{*}{M+A\_Log}  & -  & - & 0 & 1920\\
\hline
\multirow{1}{*}{M+A\_Exp}  & -  & - & 0 & 2049\\ 
\hline
\multirow{1}{*}{Forwarding}  & -  & - & 0 & 1\\ 
\hline
\end{tabular}
}
\end{table}

\begin{figure}[t]
\centering
\subfigure[Log estimation]{
\hspace{-1em}
{\scalebox{0.45}{
\includegraphics[width=\linewidth]{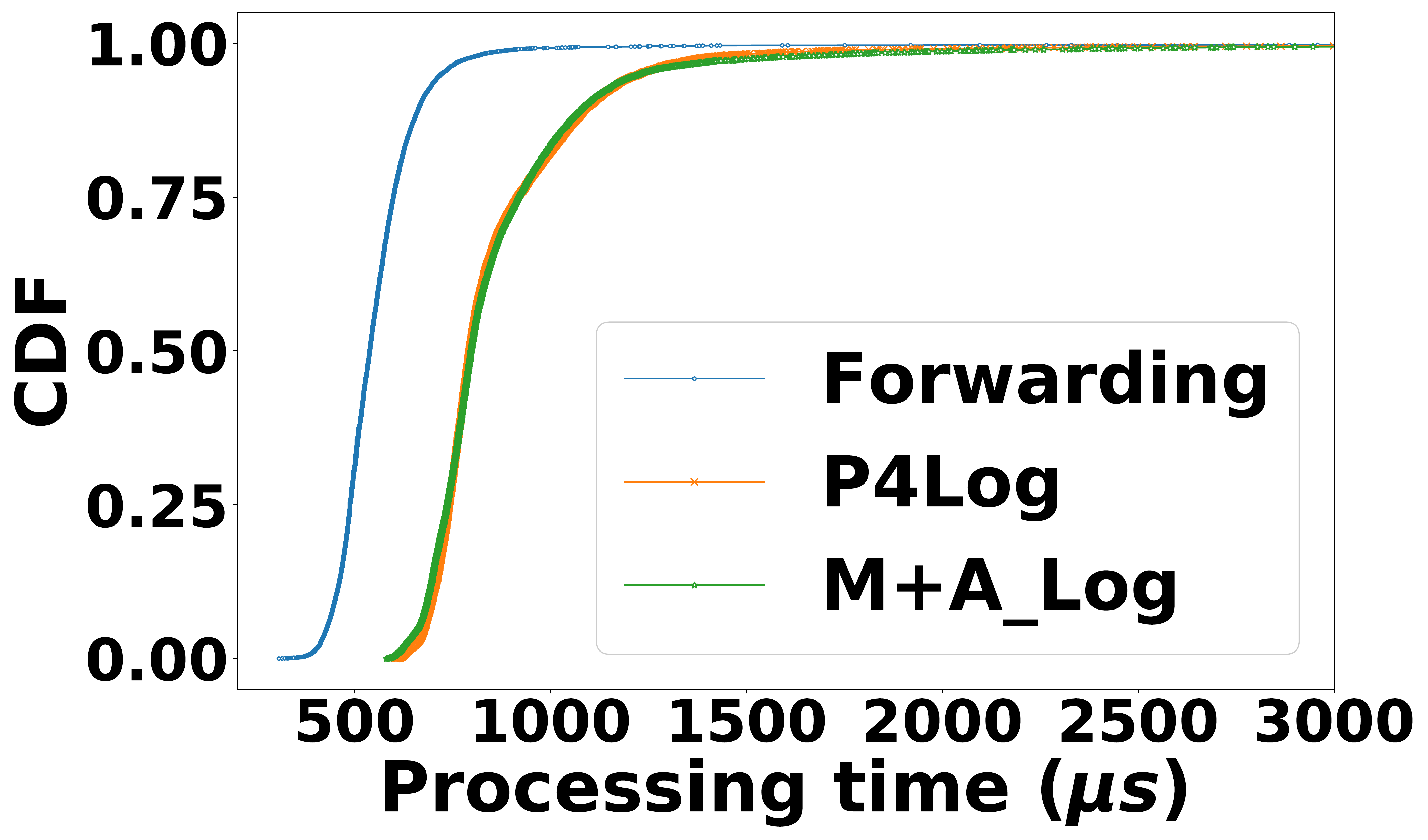}
    \label{fig:sub:cdf:p4log}
    }
    }
    }
\subfigure[Exp estimation]{
{\scalebox{0.45}{
\includegraphics[width=\linewidth]{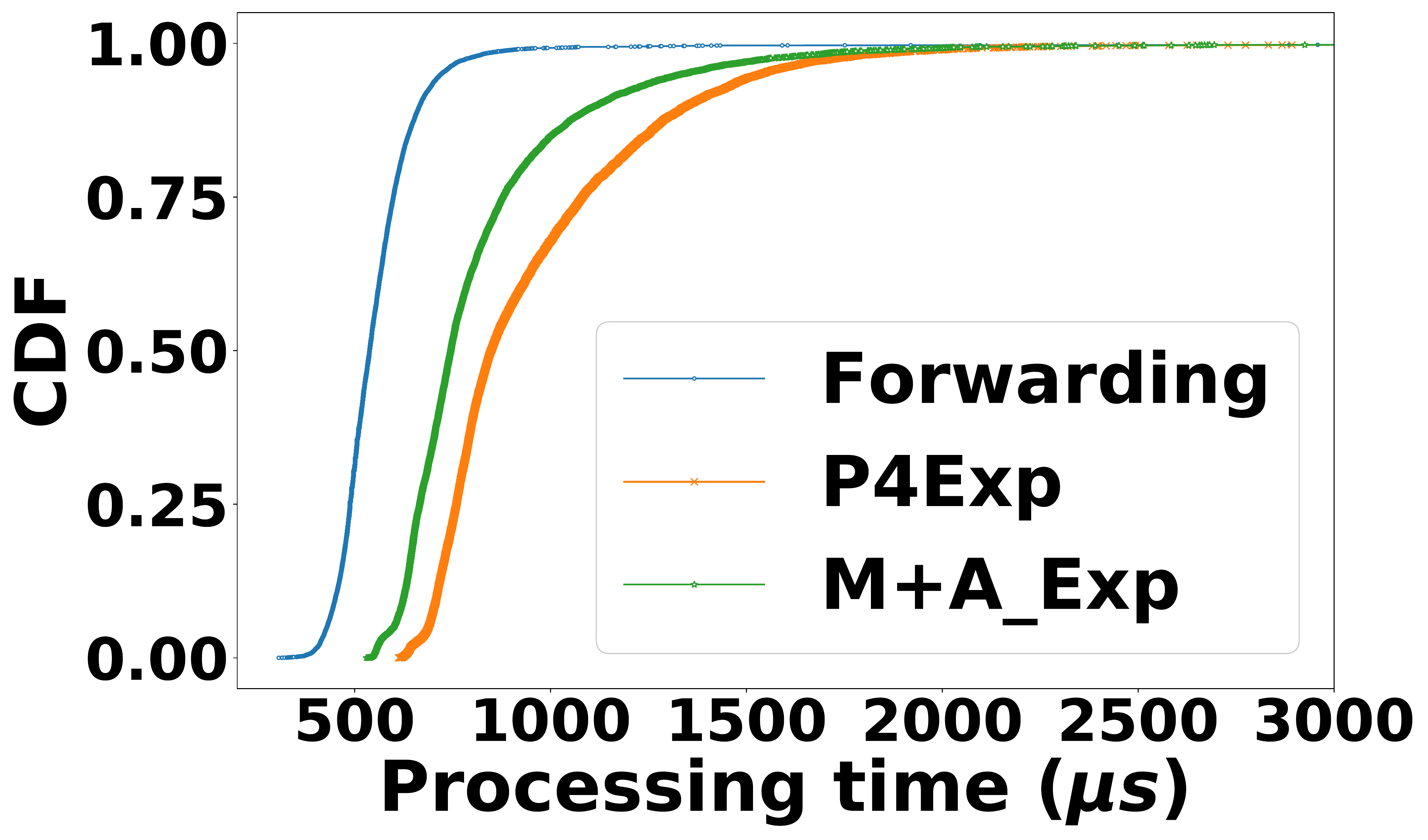}
    \label{fig:sub:cdf:p4exp}
    }
    }
    }
\caption{Cumulative distribution function of packet processing time}
\label{cdfcomparision}
\end{figure}

 Since P4 language does not support logarithm and exponential function computations, many advanced algorithms leveraging those operations (e.g., linear counting \cite{flajolet2007hyperloglog}) are not directly implementable using such domain-specific language. However, these advanced algorithms are useful for executing many networking tasks in programmable data planes, including
 flow cardinality estimation \cite{hyperloglogP4} and DDoS detection \cite{lapolli2019offloading}\cite{wang2018skyshield}, so finding a way to support them is of paramount importance. Sharma \textit{et al.} \cite{hyperloglogP4} successfully implemented estimation of logarithm and exponential function in P4, but their strategy requires the storage of appropriate pre-computed values in match-action (M+A) tables. 
 The conference version of this paper \cite{ding2019estimating} proposes and evaluates the accuracy of two algorithms for exponential function and logarithm estimation, called \emph{P4Exp} and \emph{P4Log}, which only rely on P4-supported arithmetic operations. P4Exp and P4Log algorithms have comparable accuracy as \cite{hyperloglogP4} without the usage of any M+A table.
 However, a comparison of performance in terms of \emph{packet processing time} between P4Exp, P4Log and the corresponding state of the art strategies is missing in \cite{ding2019estimating}. We believe that such a comparison is important to understand how different approaches affect packet processing in the P4 pipeline, and to take a decision on which exponential function and logarithm estimation algorithms we should leverage for the design and implementation of P4LogLog, P4NEntropy, and P4DDoS. 
 
 We chose Mininet \cite{mininet} as an emulated network environment with a single P4 switch. The data plane pipeline is described by P4 code compiled using the \emph{bmv2} behavioral model \cite{p4simulator}. We then connected the P4 switch to two hosts, ensuring that packets can be forwarded from one to the other. Virtual links bandwidth is bounded by CPU capacity. The emulated environment is built on top of a virtual machine deployed by OpenStack on our local testbed with dedicated access to 4 $\times$ 2.7GHz CPU cores and to 4GB of RAM. We used Wireshark \cite{wireshark} to capture a packet timestamp $t_{in}$ at the ingress interface of the switch and the timestamp $t_{out}$ at the egress interface when the same packet is forwarded to the destination host. The packet processing time is then calculated by  $t_{out} - t_{in}$. 
 
 In addition to P4Exp and P4Log implementations, we implemented the logarithmic and exponential function estimation strategies reported in \cite{hyperloglogP4}, named here as \emph{M+A\_Log} and \emph{M+A\_Exp}, respectively. For 64-bit operands in P4, to ensure a relative error below 1\% for the estimated values with respect to the real ones, M+A\_Exp needs 2048 entries in an exact match table, while M+A\_Log requires 1919 entries in a ternary match table (to be stored in TCAM). However, in behavioral model \cite{p4simulator}, any M+A table can include at most 1024 entries, so we had to assign two exact match tables for M+A\_Exp and two ternary match tables for M+A\_Log. A simpler benchmark strategy, named \emph{Forwarding}, is also implemented: it only requires a M+A table for forwarding the packets from source to destination host according to pre-defined flow rules. All the other strategies implement the same forwarding logic in their pipeline.
 P4Log and P4Exp parameters are taken from \cite{ding2019estimating} and shown in Tab. \ref{tuning}. The table also summarizes the number of required instructions (i.e., logical and arithmetical operations) and of M+A entries for all the considered strategies (including forwarding capabilities), showing the inherent differences of the approaches. 
 
 We then evaluate the packet processing time, considering base-2 logarithm and exponential function. We generated and forwarded 10000 packets: Fig. \ref{cdfcomparision} shows the cumulative distribution function (CDF) of packet processing time. No packet loss was experienced. As shown in Fig. \ref{fig:sub:cdf:p4log}, both P4Log and M+A\_Log cause a higher processing time than Forwarding since they need to carry out more complex operations. However, their CDF curves are almost overlapped: this means that 
 P4Log does not cause any additional overhead on processing time with respect to M+A\_Log, but it has the benefits of not requiring any M+A table.
 Likewise, Fig. \ref{fig:sub:cdf:p4exp} reveals that both exponential function estimation strategies slightly increase the processing time with respect to Forwarding. 
P4Exp has just a slightly higher packet processing time than M+A\_Exp but, also in this case, it does not require any M+A table to work. Note also that packets, in real high-performance programmable switches, are expected to be processed in few hundreds of \textit{ns} \cite{ding2021ivest}, thus such a difference in processing time would impact even less on performance (in absolute terms).  

We have shown that P4Log and P4Exp have comparable accuracy (see \cite{ding2019estimating}) and efficiency as the state of the art, while preventing from the usage of expensive and power-hungry switch memory (e.g. TCAM) for their execution: we thus chose to leverage P4Log and P4Exp for all the logarithmic and exponential-function estimations needed in the following Sections. P4Log and P4Exp can then be seen as two primitives, while the reader interested in their implementation should refer to \cite{ding2019estimating}.

As a final remark, we want to remind that, according to \cite{ding2019estimating}, the output of P4Log($x$) is $\log_{2}{x} \ll 10$ (i.e., $\log_{2}{x}$ left-shifted 10 bits), while the the output of P4Exp($x$) is $2^{x}$. 10-bit left-shifting operation $\ll$ (i.e., multiplying by $2^{10}$) is done in \cite{ding2019estimating} to "amplify" decimal numbers and maintain the information carried by their decimal part (which would be truncated by P4 otherwise). 10-bit right-shifting $\gg$ can be instead used to move back an amplified number to its original range. Note that P4Exp can be used for the exponential function computation of any real positive number, while in the case of natural numbers it is more efficient to exploit the left shift operator (i.e., $2^x \leftarrow 1\ll x$ for $x$ integer positive). We will largely leverage such properties and logical operations in the next sections of this paper.

\section{Estimation of normalized traffic entropy} \label{sec:nentropy}

Based on P4Log and P4Exp we initially propose \emph{P4LogLog} to estimate flow cardinality. P4LogLog is then used by \emph{P4NEntropy}, which estimates the normalized network traffic entropy. 
The prototypes of both strategies have been implemented in P4 behavioral model \cite{p4simulator} and are executable in an emulated environment as Mininet \cite{mininet}. The P4 source codes are available in \cite{P4LogLog} and \cite{P4NEntropy}. 

\subsection{Flow cardinality estimation: P4LogLog} \label{sec:P4loglog}

In this section we propose P4LogLog for the estimation of flow cardinality. 
The problem is formulated as follows.

\textbf{Problem definition:} \emph{Given} a stream $S$ of incoming packets, each one belonging to a specific \emph{flow} $i$, \emph{returns} the estimated \emph{flow cardinality} $\hat{n}$ of $S$, i.e., the estimated number of distinct flows in $S$. 

For instance, if we identify as \emph{flow key} $i$ each packet \emph{destination IP}, meaning that a flow includes all the packets towards a specific destination, then the flow cardinality of destination IPs represents the number of destination IPs in the network. Same consideration holds for any other flow definition (e.g. packets with the same 5-tuple, same source/destination IP pair, etc.) without any loss of generality. In the following, we report the details of Update and Query operations of P4LogLog, which both follow specifications from LogLog \cite{durand2003loglog} (see Section \ref{background:distinct}) while only using P4-supported instructions.  

\subsubsection{Update}
As shown in Algorithm \ref{P4LogLogU}, \emph{Update} function iteratively updates a readable and writable stateful register $Reg$ for each incoming packet, which belongs to a flow with flow key $i$. The flow key $i$ of the packet is hashed by a given hash function, and the output value is converted to a $os$-bit binary string $s$ (Line 6). In this paper, we consider $os=32$ and an $m$-sized register $Reg$, where $m=2^k$ and integer $k \in \{4, ...,  16\}$ (as per \cite{durand2003loglog}). The index of the register's cell to be updated, named $bucket$ ($ 0 \le bucket \le m-1$), is the binary number represented by the rightmost $k$ bits of $s$, which can be obtained by $s \& (2^{k}-1)$, i.e., $s \& 0 \underbrace{11 \cdots 1}_{k}$ (Line 7). \& is the bitwise inclusive AND operator and $2^{k}-1$ (in binary) is pre-stored in the P4 program once $k$ is chosen. The algorithm then right-shifts $s$ to $k$ bits to get a binary string $x$ where the first $k$ bits are 0s and the remaining $os-k$ bits are the first $os-k$ bits of $s$ (Line 8). The index of rightmost 1 in $x$, called \emph{value}, is then used to update the LogLog register's cell in $bucket$ position. Unfortunately, retrieving such rightmost 1 is not trivial. As shown from Lines 9 to 12, the algorithm adopts the following strategy: all bits of $x$ on the left of the rightmost 1 are iteratively converted to 1, and the result of this iterative operation is stored in $w$ ($\: |$ is the bitwise inclusive OR operator). For example, an $os$-bit binary value $x=\underbrace{00 \cdots0 1}_{os-1}0$ is converted to $w=\underbrace{11 \cdots11}_{os-1}0$. The algorithm for \emph{Hamming weight} recalled in Section \ref{hamming} is then used to count $b$, i.e., the number of 1s in $x$ (Line 12): $value$ is equal to $os + 1 - b$ (Line 13). 
Finally, if $value$ is larger than the $bucket$-indexed value in the register, $value$ replaces the stored value (Lines 14-15). 
\begin{algorithm} [t]
\caption{\textbf{P4LogLog}}
\label{P4LogLogU} 
\small
 \SetKwProg{Fn}{Function}{:}{}

  \KwIn{Packet stream $S$} 

  \KwOut{Flow cardinality estimation $\hat{n}$} 
  
 $m \leftarrow 2^{k}$ $(k \in \{4, ...,  16\})$\\
 $os \leftarrow 32$\\
 
\emph{Reg} $\leftarrow$ $m$-sized empty LogLog register\\

\Fn{Update$(Reg)$}{
\For{Each received packet belonging to flow $i$}{
    $s \leftarrow (Hash(i) \rightarrow \{0, 1\} ^ {os} )$\\
	$bucket \leftarrow s \& (2^{k}-1)$\\
	$x \leftarrow (s \gg k )$\\
    $w \leftarrow x  | (x\ll1)$\\
    \For {int $l$ $\in \{1, \cdots,\log_2(os)-1\}$}{
        $w \leftarrow w  | (w \ll 2^{l})$\\
    }
		$b \leftarrow$ $HammingWeight(w)$ \\
	$value \leftarrow os + 1 - b$\\
\If{$value > Reg[bucket]$}{
		$Reg[bucket] \leftarrow value$
}
}
\Return{$Reg$}
}
$\alpha_{m} \leftarrow 0.39701 \ll 10$\\
\Fn{Query$(Reg)$}{
 $exp \leftarrow P4Exp((\sum_{bucket=0}^{m-1}{ Reg{[bucket]} }) \gg k) $\\
 $\hat{n} \leftarrow  (exp\cdot \alpha_{m} \cdot m)\gg 10  $\\
\Return{$\hat{n}$}
}

\end{algorithm}

\subsubsection{Query}
Query function in Algorithm \ref{P4LogLogU} estimates the flow cardinality directly in the switch.
The flow cardinality estimation $\hat{n}$ is computed as in \cite{durand2003loglog} and Section \ref{background:distinct} from all LogLog register's stored values by exploiting P4Exp. The $k$-bit right-shift operation carried out on the sum of values from $Reg$ is equivalent to dividing such sum by $m=2^k$ (Line 19). The floating parameter $\alpha_{m}$, chosen as in \cite{durand2003loglog}, is amplified $2^{10}$ times through left shift operation, and the resulted value from the computation executed in Line 20 is right-shifted 10 bits to get the estimated flow cardinality $\hat{n}$. 

%



\subsection{Normalized traffic entropy estimation: P4NEntropy}
 In this section we present a new strategy, named \emph{P4NEntropy}, to  estimate the normalized network traffic entropy in a given time interval using the P4 language. 
Formally, the problem is defined as follows.

\textbf{Problem definition:} \emph{Given} a stream $S$ of incoming packets, each one belonging to a specific \emph{flow} $i$, and a time interval $T_{int}$, \emph{returns} the \emph{normalized Shannon entropy} estimation $H_{norm}$ (see Section \ref{sec:entropy}) at the end of $T_{int}$. 

\subsubsection{Derivation of estimated normalized entropy in P4} \label{sec:entropy_derivation}
The goal of this section is to provide an estimation of network traffic normalized entropy by only using P4-supported operations and reducing as much as possible their number. 
The section also shows how relevant statistics, used for normalized entropy estimation at the end of $T_{int}$, are iteratively updated every time a packet crosses the switch. 

We first rewrite the Shannon entropy as follows:

{\small
\begin{align*}
		H(|S|_{tot}) &= -\sum_{i=1}^{n}{\frac{f_i(|S|_{tot})}{|S|_{tot}} \log_{d} {\frac{f_i(|S|_{tot})}{|S|_{tot}}}}\\
		  &=\log_{d}|S|_{tot}- \frac{1}{|S|_{tot}} \sum^{n}_{i=1}{f_i(|S|_{tot})\log_{d}{f_i(|S|_{tot})}}
\end{align*}
}

\noindent We consider $d=2$ without any loss of generality. 
With respect to the definition given in Section \ref{sec:entropy}, we use the notation $f_i{(|S|_{tot})}$ to make explicit that $f_i$ refers to its value when $|S|_{tot}$ packets have been received (i.e., at the end of $T_{int}$). 
As packets arrive to the switch, the overall number of processed packets $|S|$ increases and must be stored in the switch to ensure that $H(|S|_{tot})$ can be computed at the end of $T_{int}$, when $|S|=|S|_{tot}$. We define $Sum(|S|)=\sum^{n}_{i=1}~{f_i(|S|)\log_{d}{f_i(|S|)}}$, which must be updated as well. To understand how to update $Sum(|S|)$, let's assume that a new packet for a specific flow arrives and it is the $|S|$-th packet. We call its packet count $\bar{f_i}(|S|)$. It holds that:

{\small
\begin{equation*} 
    	\begin{cases}  f_i(|S|)=f_i(|S|-1) \hspace{30pt} (f_i(|S|) \neq  \bar{f_i}(|S|)) \\ f_i(|S|)=f_i(|S|-1) +1  \hfill (f_i(|S|) =  \bar{f_i}(|S|)) \end{cases}
\end{equation*}
}

\noindent This allows us to re-write $Sum(|S|)$ as follows:

{\small
\begin{align*} 
    Sum(|S|) = \hspace{4pt} &Sum(|S|-1) + \bar{f_i}(|S|)\log_2\bar{f_i}(|S|) \hspace{2pt}+\\\nonumber&-(\bar{f_i}(|S|)-1)\log_2(\bar{f_i}(|S|)-1)
\end{align*}}

\noindent $Sum(|S|)$ thus needs two logarithmic computations for each incoming packet, and would require running P4Log twice with corresponding computational effort. 

In the next step, we show how it is possible to estimate $Sum(|S|)$ with only (at most) one logarithmic computation. 
When $\bar{f_i}(|S|)=1$, we estimate $Sum(|S|)=Sum(|S|-1)$, being $\bar{f_i}(|S|)\log_2\bar{f_i}(|S|)=1\log_2 1 =0$ and defining $(\bar{f_i}(|S|)-1)\log_2(\bar{f_i}(|S|)-1)=0\log_2 0 = 0$ \cite{liu2016one}. Instead, when $\bar{f_i}(|S|)>1$, we need to re-write once again $Sum(|S|)$ in the following way:

{\small
\begin{align*} 
    \nonumber Sum(|S|)= \hspace{4pt} &Sum(|S|-1) + \log_2\bar{f_i}(|S|)\hspace{2pt}+\\\nonumber&+(\bar{f_i}(|S|)-1)\log_2 (1+\frac{1}{\bar{f_i}(|S|)-1})
\end{align*}
}

\noindent According to L'Hopital's rule \cite{struik1963origin}:

{\small
\begin{align*}
\lim _{\bar{f_{i}}(|S|) \rightarrow +\infty}{(\bar{f_{i}}(|S| - 1)\log_2{(1 + \frac{1}{(\bar{f_{i}}(|S| - 1)})}} = \frac{1}{\text{ln}2}
\end{align*}}

 \noindent Thus, we set $1/\text{ln}2\approx 1.44$ as the approximation of the third term of $Sum(|S|)$. This approximation best works when most of the flows in $T_{int}$ carry a number of packets much greater than 1  (as it usually happens in an ISP backbone network, which is the most suitable scenario where to apply our strategy).
Finally, $Sum(|S|)$ can be estimated as:

{\small
\begin{align} \label{eq1}
    	 Sum(|S|) \approx &\begin{cases}  Sum(|S|-1) \quad\quad\quad\quad (\bar{f_{i}}(|S|) = 1) \\ \nonumber Sum(|S|-1) + \log_2{\bar{f_{i}}(|S|) } + 1/\text{ln}2\vspace{-1pt}\end{cases} \\ &\quad\quad\quad\quad\quad\quad\quad\quad\quad\quad\quad(\bar{f_{i}}(|S|) > 1)
\end{align}}

\noindent This estimation requires at most one logarithm computation. Since P4 language does not support division, we re-write $\frac{1}{|S|_{tot}}=2^{-\log_2{|S|_{tot}}}$. So, entropy can be written as:

{\small
\begin{align*}
H(|S|_{tot}) = \log_2|S|_{tot}- 2^{(\log_2{Sum(|S|_{tot})} - \log_2{|S|_{tot})}}
\end{align*}}

\noindent In this form, entropy can be estimated by only using P4-supported operations, leveraging P4Log and P4Exp algorithms. In the following we show how, in some cases, it is possible to further slightly reduce complexity in entropy estimation.
When $|S|_{tot}=\sum_{i=1}^{n}{f_{i}(|S|_{tot})}>Sum(f_{i}|S_{tot}|)$, it holds that $0 < 2^{(\log_2{Sum(|S|_{tot})} - \log_2{|S|_{tot})}} < 1$. This is a corner case that happens only when flow distribution is almost uniform (i.e., when most of flows carry only one or very few packets). In this case, we neglect the computation of $2^{(\log_2{Sum(|S|_{tot})} - \log_2{|S|_{tot})}}$, meaning that we estimate entropy as flow distribution was perfectly uniform. Network traffic entropy can then be estimated as follows:

{\small
\begin{align} \label{eq2}
    	\nonumber H(|S|_{tot}) \approx &\begin{cases}  \log_2(|S|_{tot}) \quad\quad(|S|_{tot} > Sum(|S|_{tot})) \\\log_2(|S|_{tot}) - 2^{(\log_2{Sum(|S|_{tot})} - \log_2{|S|_{tot})}}
    	\vspace{-4pt}\end{cases}\\&\quad\quad\quad\quad\quad\quad\quad\quad (|S|_{tot} \le Sum(|S|_{tot}))
\end{align}}

\noindent Finally, normalized entropy $H_{norm}(|S|_{tot})$ is estimated as: 

{\small
\begin{align}\label{eq3}
    H_{norm}(|S|_{tot}) = 2^{\log_2(H(|S|_{tot}))-\log_2(\log_2{ \hat{n}})}
\end{align}}

\noindent  The number of estimated distinct flows $\hat{n}$ can be obtained using P4LogLog, that is, by updating a LogLog register for each incoming packet and by querying it at the end of $T_{int}$.

 \begin {figure}[t]
\centering
\scalebox{0.5}{
\begin{tikzpicture}[]
\node at  (-1.5,5) [draw,circle,align=center] (stream) {\Large{Packet} \\ \Large{stream $S$}};


\node  (hll3) at  (3.5,5) [draw,thick,minimum width=2cm,minimum height=1cm,align=center] { \Large{Update} \\ \Large{counter} \\ \Large{$|S|$}};


\node  (ll_dst) at  (3.5,10) [draw,thick,minimum width=2cm,minimum height=1cm,align=center] { \Large{P4LogLog$(S)$}};

\draw [->, thick] (3.5, 5.8) --  (3.5, 7.5) node [midway, left,align=center]  (TextNode) {\Large{At the end of $T_{int}$}\\ \Large{$(|S| = |S|_{tot})$}};

\draw [->, thick] (stream) --  (hll3) node [midway, above, sloped]  (TextNode) {$dstIP_i$};
\draw [->, thick] (stream) |-  (ll_dst) node [pos=.85,  above, sloped]  (TextNode) {$dstIP_i$};



\node (fastCM) at  (8, 5)[rectangle, draw, inner sep=0]{
\mygrid
};
\node  (kk) at  (8,6.5) [minimum width=2cm,minimum height=1cm] { \Large{Update / Query Sketch}};
\draw [->, thick] (hll3) --  (fastCM)node [midway, above, sloped]  (TextNode) {};

\node  (samplelist-mid) at  (14,5) [draw,thick,minimum width=3cm,minimum height=1.9cm, align=center] {\Large{Update}\\  \Large{$Sum(|S|)$} \\\Large{(Eq. \ref{eq1})}};

\draw [->, thick] (fastCM) --  (samplelist-mid)node [right=0.1cm, above, midway]  (TextNode) {$\bar{f_{i}}(|S|)$};

\node  (thresholdHH) at  (3,8) [draw,thick,minimum width=2cm,minimum height=1cm] {\Large{Estimate Entropy $H(|S|_{tot})$} (Eq. \ref{eq2})};
\draw [<-, thick](thresholdHH) -|  (samplelist-mid)node [pos=.3, above, align =center]  (TextNode){\Large{At the end of $T_{int}$}\\ \Large{$(Sum(|S|) = Sum(|S|_{tot}))$}};
\node  (norm) at  (12,10) [draw,thick,minimum width=2cm,minimum height=1cm, align =center] {\Large{Estimate Norm. Entropy }\\ \Large{$H_{norm}(|S|_{tot})$ (Eq. \ref{eq3})}};
\draw [->, thick] (ll_dst) --  (norm) node[midway,align=center, above]  (TextNode) {\Large{At the end of $T_{int}$}\\\Large{$(\hat{n}_{dst})$}};;
\draw [-, thick] (6,8.5) --  (6 ,10);
\end{tikzpicture}
 }
\caption{Scheme of P4NEntropy}
\label{nwh}
\end{figure}
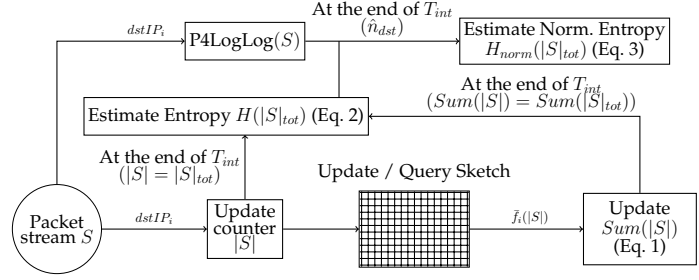

\begin{algorithm} [t]
\caption{\textbf{P4NEntropy}}
\label{alg:entropy} 
\small
 \SetKwProg{Fn}{Function}{:}{}
\KwIn{Packet stream $S$, time interval $T_{int}$}
\KwOut{Normalized entropy estimation $H_{norm}(|S|_{tot})$ of $S$ in $T_{int}$} 
  $|S| \leftarrow 0$, $Sum{(|S|)} \leftarrow 0$\\
\Fn{UpdateSum$(Sum(|S|))$}{
  \While {currentTime $<$ $T_{int}$}{
  		\For {Each received packet belonging to flow $i$}{
  		    $|S| \leftarrow |S| + 1$\\
  		    $\bar{f_i}(|S|) \leftarrow Sketch({dstIP}_i)$\\

\If{$\bar{f_i}(|S|) > 1$}{
$Sum(|S|) \ll 10 \leftarrow Sum(|S|)\ll 10$\\ $+ P4Log(\bar{f_i}(|S|)) + 1.44\ll10$
}
 		}	
 		}
 $Sum(|S|_{tot}) \leftarrow (Sum(|S_{tot}|) \ll 10)\gg 10$\\
 \Return $Sum(|S|_{tot})$, $|S|_{tot}$

}

\Fn{EstimateNormEntropy($Sum{(|S|_{tot})}$, $|S|_{tot}$)}{
\If {currentTime = $T_{int}$}{
   \If{$|S|_{tot} > Sum(|S|_{tot})$}{\hspace{-8pt}$H(|S|_{tot}) \ll 10 \leftarrow P4Log(|S|_{tot})$}
        
   \Else{
   \hspace{-9pt} $\text{diff} \leftarrow P4Log(Sum(|S|_{tot}))-P4Log(|S|_{tot})$\\
   \hspace{-8pt}$H(|S|_{tot}) \hspace{-2pt} \ll \hspace{-2pt} 10 \hspace{-2pt} \leftarrow \hspace{-3pt} P4Log(|S|_{tot})- P4Exp(2,\text{diff})$
   }
   $\hat{n}_{dst} \leftarrow P4LogLog(S,T_{int})$\\
  $\text{diff}_{n} \leftarrow P4Log(H(|S|_{tot}) \ll 10)+$\\$-(P4Log(P4Log(\hat{n}_{dst}))-10\ll10)$ \\
  \If{$\text{diff}_{n} > 0$}{
  $H_{norm}(|S|_{tot}) \ll 10 \leftarrow P4Exp(2,\text{diff}_{n})$
  }\Else{
    $H_{norm}(|S|_{tot}) \ll 10  \leftarrow  0$
  }
}
\Return $H_{norm}(|S|_{tot}) \ll 10$
}
\end{algorithm}

\subsubsection{Description of P4NEntropy strategy}
Figure \ref{nwh} and Algorithm \ref{alg:entropy} show the scheme and pseudocode of P4NEntropy algorithm, leveraging outcomes from Sections \ref{sec:P4loglog} and \ref{sec:entropy_derivation}. First, the algorithm continuously updates $Sum(|S|)$ until the end of $T_{int}$ (\emph{UpdateSum} function) with flow information from incoming packets. A counter $|S|$ is used to count all incoming packets in the switch. 
Note that we consider as \emph{flow key} the destination IP of the packet, with $i\sim dstIP_i$. 
A sketch data structure (e.g., Count Sketch or Count-min Sketch, see Section  \ref{sec:background_sketch}) is used to store the estimated packet count for all the flows, being continuously updated to include information from new packets, and then it is queried to retrieve the estimated packet count $\bar{f_{i}}(|S|)$ for the flow  $i$ the current incoming packet belongs to. This value is then passed to a register named $Sum{(|S|)}$, which is updated as specified in Eq. \ref{eq1}. All the floating numbers in the equation must be amplified $2^{10}$ times, since P4Log outputs an amplified integer value. Only at the end of $T_{int}$, $Sum(|S|_{tot})$ is reduced by a factor of $2^{10}$ and its final value, together with $|S|_{tot}$, is returned (Lines 1-11 of the pseudocode).
Traffic entropy is then estimated as specified in Eq. \ref{eq2} (Lines 13-18). The resulted value of $H(|S|_{tot})$ is amplified $2^{10}$ times since output values of P4Log are amplified, while output values of P4Exp are not. Such an amplification makes it possible to use P4Exp in Eq. \ref{eq3} to estimate $H_{norm}(|S|_{tot})$ amplified $2^{10}$ times. 
Note that $H(|S|_{tot}) \ll 10$ may be smaller than $\log_2(\hat{n}_{dst})$
but, in this case, the normalized network traffic entropy can be approximated to 0 (Line 25). Since the result of P4Log is left-shifted 10 bits, the computation of $\log_2(\log_2(\hat{n}_{dst}))$ must be carefully handled.  Considering that the result of P4Log($\hat{n}_{dst}$) is $\log_2(\hat{n}_{dst}) \ll 10$,  the output of P4Log($\log_2(\hat{n}_{dst}) \ll 10$) can be expressed as $\log_2(\log_2(\hat{n}_{dst}) \ll 10) \ll 10 = \log_2(\log_2(\hat{n}_{dst}) \cdot 2^{10}) \ll 10 = \log_2(\log_2(\hat{n}_{dst})) \ll 10 + 10 \ll 10$. Hence, $\log_2(\log_2(\hat{n}_{dst})) \ll 10$ is equivalent to P4Log(P4Log($\hat{n}_{dst}$))  $ - 10 \ll 10$ (Line 21).
The resulting value is used to compute the normalized network traffic entropy amplified $2^{10}$ times (Line 23).

\vspace{-0.5em}
\section{Entropy-based DDoS detection} \label{sec:ddos}
Based on P4NEntropy, we present a simple yet effective entropy-based DDoS detection strategy in P4, named \emph{P4DDoS}. The P4 code of P4DDoS is available in \cite{P4DDoS}. Formally, the problem is defined as follows.

\textbf{Problem definition:} \emph{Given} a $k$-th time interval $T_{int}^{k}$, a stream $S_k$ of incoming packets during $T^k_{int}$, the estimated normalized network traffic entropy of destination IPs $H^k_{norm}$ at the end of $T^k_{int}$ and an adaptive threshold $\lambda_{norm}^{k}$, \emph{returns} an \emph{alarm} to the controller, at the end of $T^k_{int}$, if a \emph{potential DDoS attack} is identified. 

Our proposed strategy triggers an alarm (e.g. a flag embedded in a field of the \textit{report packet} header) if $H^k_{norm} < \lambda_{norm}^{k}$. In fact, as empirically evaluated in previous works (e.g. \cite{entropyDDoS}\cite{afek2018detecting}), when a DDoS attack occurs, the normalized network traffic entropy of destination IPs significantly decreases, since traffic is concentrated around few destination nodes. The most critical aspect for such an entropy-based strategy is how to set the threshold $\lambda_{norm}^{k}$. This will be discussed in the next subsection.
Note also that we only focus on \textit{volumetric} DDoS attacks (e.g. UDP flooding or DNS amplification attacks); considering other types of attacks, such as \textit{link flooding} or \textit{carpet bombing}, is left as future work.
\subsection{Adaptive threshold setting} \label{subsec:threshold}
Since network traffic fluctuates over time, we define an \emph{adaptive threshold} to protect our strategy from false positives that may be generated if using a fixed-value threshold in such a dynamic environment. 
Our proposed adaptive threshold leverages the computation of an \emph{Exponentially Weighted Moving Average} (EWMA) of $H^k_{norm}$ across different time intervals.
The moving average $EWMA^k_{norm}$ in time interval $T^k_{int}$ is expressed as:

{\small
\begin{align*}
    EWMA_{norm}^{k} = \begin{cases}
    H_{norm}^{k} \quad\quad\quad\quad\quad\quad\quad\quad\quad\quad\quad\; (k = 1) \\
    \alpha H_{norm}^{k} + (1 - \alpha) EWMA_{norm}^{k-1}  (k > 1)
    \end{cases}
\end{align*}}

where $\alpha \quad (0<\alpha<1)$ is the smoothing factor for $EWMA^k_{norm}$. We define a threshold parameter $\epsilon$ $(0\le \epsilon \le 1)$, used to compute the threshold $\lambda_{norm}^{k+1}$ in the next time interval $T^{k+1}_{int}$ if no alarm is generated in $T^{k}_{int}$:

{\small
\begin{align*}
    \lambda_{norm}^{k+1} = \begin{cases}
    EWMA_{norm}^{k} - \epsilon \quad\quad\quad\text{(no alarm in $T^k_{int}$)} \\
    \lambda_{norm}^{k} \quad\quad\quad\quad\quad\quad\quad\quad \text{(alarm in $T^k_{int}$})
    \end{cases}
\end{align*}}

As shown above, the threshold $\lambda_{norm}^{k+1}$ is not updated if an alarm is generated in the time interval: this ensures that the threshold is updated when only legitimate traffic crosses the switch and its value is not biased by DDoS traffic. Note that setting the parameter $\epsilon$ in a proper way is also fundamental to get good DDoS detection performance. This aspect will be evaluated in Section \ref{sec:evaluation2}.


\subsection{Implementation in P4 language}
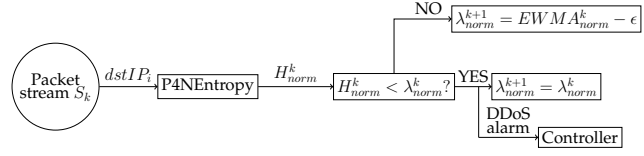
\begin {figure}[t]
\centering
\scalebox{0.45}{
\begin{tikzpicture}[]
\node at  (-1.5,8) [draw,circle,align=center] (stream) {\Large{Packet} \\ \Large{stream $S_k$}};
\node at  (3,8) [draw,align=center,minimum width=2cm] (P4NEntropy) {\Large{P4NEntropy}};
\node at  (8.5,8) [draw,align=center,minimum width=2cm] (Tnorm) {\Large{$ H_{norm}^{k}< \lambda_{norm}^{k}?$}};
\node at  (13,8) [draw,align=center,minimum width=2cm] (norm_true) {\Large{$\lambda_{norm}^{k+1} = \lambda_{norm}^{k}$}};
\node at  (13,10) [draw,align=center,minimum width=2cm] (norm_false) {\Large{$\lambda_{norm}^{k+1}=EWMA_{norm}^{k} - \epsilon$}};
\node at  (14,6.5) [draw,align=center,minimum width=2cm] (controller) {\Large{Controller}};

\draw[->, thick](stream) -- (P4NEntropy) node [above,pos=0.5,align=center]  (TextNode) {\Large{$dstIP_{i}$}};
\draw[->, thick](P4NEntropy) -- (Tnorm) node [above,midway,align=center]  (TextNode) {\Large{$H_{norm}^{k}$}};
\draw[->, thick](Tnorm) -- (norm_true) node [above,midway,align=center]  (TextNode) {\Large{YES}};
\draw[->, thick](Tnorm) |- (norm_false) node [above,pos=0.8,align=center]  (TextNode) {\Large{NO}};

\draw[-, thick](11, 8) -- (11, 6.5);
\draw[->, thick](11, 6.5) -- (controller)node [above,midway,align=center]  (TextNode) {\Large{DDoS}\\ \Large{alarm}};

\end{tikzpicture}
 }
\caption{Scheme of P4DDoS}
\label{ddosscheme}
\end{figure}
\begin{algorithm} [t]
\caption{\textbf{P4DDoS}}
\label{alg:ddos} 
\small
 \SetKwProg{Fn}{Function}{:}{}
\KwIn{Packet stream $S_k$, time interval $T_{int}^{k}$, threshold parameter $\epsilon$, smoothing factor $\alpha$, threshold $\lambda_{norm}^k\ll 10$ and average $EWMA_{norm}^{k-1}\ll 10$ computed in $T_{int}^{k-1}$}
\KwOut{DDoS alarm $Alarm_{ddos}^{k}=1$ if a DDoS attack is detected in $T_{int}^{k}$}
\Fn{DDoSDetection$(\lambda_{norm}^k\ll 10)$}{
    $H_{norm}^{k} \ll 10 \leftarrow$ P4NEntropy$(S_k,T^k_{int})$\\ 
    $Alarm_{ddos}^{k} \leftarrow 0$\\
    \If{$H_{norm}^{k} \ll10 < \lambda_{norm}^{k} \ll 10 $}{
            $Alarm_{ddos}^{k} \leftarrow 1$\\
            \emph{UpdateThreshold}$(k, \alpha, H_{norm}^{k}\ll 10, \epsilon, Alarm_{ddos}^{k},$\\ $EWMA_{norm}^{k-1}\ll10,\lambda_{norm}^k\ll 10)$
    }\Else{
       \emph{UpdateThreshold}$(k, \alpha, H_{norm}^{k}\ll 10, \epsilon, Alarm_{ddos}^{k},$\\ $EWMA_{norm}^{k-1}\ll 10,\lambda_{norm}^k\ll 10)$
    }
    \Return{$Alarm_{ddos}^{k}$}
}

\Fn{UpdateThreshold$(k, \alpha, H_{norm}^{k} \hspace{-0pt}\ll \hspace{-0pt}10, \epsilon,$\\$Alarm_{ddos}^{k},EWMA_{norm}^{k-1}\ll 10,\lambda_{norm}^k\ll 10)$}{
  		\If {$Alarm_{ddos}^{k}=0$}{
  		\If{ $k=1$}{
  		    $EWMA_{norm}^{k} \ll 10 \leftarrow H^k_{norm} \ll 10$
  		    }
  		\Else{
  		$EWMA_{norm}^{k} \hspace{-2pt}\ll \hspace{-2pt}10 \leftarrow ((\alpha \ll 10) \cdot  H_{norm}^{k} \hspace{-2pt}\ll 10\hspace{-2pt}$ \\$+((1-\alpha) \ll 10) \cdot EWMA_{norm}^{k-1}\ll 10)\gg10 $
  		 
  		}
        $\lambda_{norm}^{k+1} \ll 10 \leftarrow EWMA_{norm}^{k} \ll 10 - \epsilon \ll 10$
        }
        \Else{
        $\lambda_{norm}^{k+1} \ll 10 \leftarrow \lambda_{norm}^{k} \ll 10$
        }
        \Return{$\lambda_{norm}^{k+1}\ll 10,EWMA_{norm}^{k}\ll 10$}
}
\end{algorithm}

Figure \ref{ddosscheme} and Algorithm \ref{alg:ddos} report the scheme and pseudocode of the P4DDoS strategy, with focus on a given time interval $T^k_{int}$. At the end of time interval $T_{int}^{k}$, the \emph{DDoSDetection} function is executed. $Alarm_{ddos}^k$ is set to 0 and the normalized network traffic entropy $H_{norm}^{k}$ is estimated by P4NEntropy, amplified $2^{10}$ times (Lines 2-3). It is then compared to the threshold $\lambda^{k}_{norm}\ll 10$ (Line 4). If smaller, the alarm $Alarm_{ddos}^{k}$ is set to 1 and the \emph{UpdateThreshold} function is called
(Lines 5-7). 
Otherwise, the \emph{UpdateThreshold} function is called without changing $Alarm_{ddos}^{k}$ (Lines 9-10). If $Alarm_{ddos}^{k}=1$, the switch clones the current packet and embeds the value 1 in a customized header field.  This report packet is then sent to the controller to report that a \textit{potential} DDoS attack has been detected. It is possible to embed more information in the  header of the report packet, such as the estimated network traffic entropy. In this case, the controller is able to take a network-wide decision using the entropy retrieved from multiple switches (see Section \ref{sec:nw}) about whether the \textit{potential} DDoS attack is an \textit{actual} attack. 

The \emph{UpdateThreshold} function updates EMWA and the adaptive threshold as specified in Section \ref{subsec:threshold}. (Lines 12-23). Note that, since both EMWA and the threshold $\lambda_{norm}$ are usually decimal numbers, all the operations are executed to ensure that their value is amplified $2^{10}$ times.


\subsection{Insights and discussions}
\subsubsection{Network-wide coordination} \label{sec:nw}
So far, we have focused on entropy-based DDoS detection in a single programmable switch. The switch can generate alarms if, according to the traffic flowing through its interfaces, a DDoS attack may be occurring. However, given the reduced network visibility of a single switch, a final decision on whether a DDoS attack is actually carried out should be taken by the centralized controller from a \emph{network-wide} perspective, that is, by cross-checking collected information from multiple switches and taking a global decision. 
For instance, UnivMon \cite{liu2016one} and Elastic Sketch \cite{yang2018elastic} present a way to estimate \emph{network-wide traffic entropy}: the idea behind those works is to sample a set of flows with large packet count in any programmable switch, and send such statistics to the controller at the end of any time interval. The controller estimates the entropy of reported sampled "heavy" flows and considers it as a \emph{network-wide entropy estimation}. As reported in \cite{basat2018network}, these two approaches assume that packets for a specific flow are counted only once in the network. By making the same strong assumption, 
in our case network-wide traffic entropy $H^{nw}$ can be expressed as:

{\small
\begin{align*}
    H^{nw} = \log_{2}(\sum_{j=1}^{w}|S_j|_{tot}) - \frac{1}{\sum_{j=1}^{w}|S_j|_{tot}}(\sum_{j=1}^{w}{Sum_{j}})
\end{align*}}

where $w$ is the number of switches in the network and $Sum_{j}= \sum^{n_j}_{i=1}~{f_i(|S_j|_{tot})\log_{d}{f_i(|S_j|_{tot})}}$ (see Section \ref{sec:entropy_derivation}). Additionally, according to the union property of LogLog (see section \ref{background:distinct}), the normalized network-wide traffic entropy $H_{norm}^{nw}$ can be expressed as:

{\small
\begin{align*}
    H_{norm}^{nw} = \frac{H^{nw}}{\log_{2}(LogLog(S_1 \cup S_2 \cup \cdots S_w))}
\end{align*}}

In this latter case, the strong assumption above can be neglected, since the union property of LogLog makes it possible to estimate the network-wide number of distinct flows also if a packet is counted in different locations.

Given the above considerations, a network-wide strategy could be designed to forward to the controller all the needed information from the switches (i.e., $|S_j|_{tot}$, $Sum_j$ and the $j$-th LogLog register) for the computation of network-wide normalized entropy in support to a centralized \emph{network-wide DDoS detection}. 
However, since in real scenarios the packet may traverse multiple switches and generate duplicated  packet counts, the accuracy of the computed network-wide entropy $N^{nw}$ would be compromised. How to overcome this issue is still open: we will work on refined strategies for network-wide entropy-based DDoS detection in the future. 
\subsubsection{Implementation in a programmable hardware switch} 
We tried to implement P4DDoS in a P4-programmable hardware switch with Tofino Application-Specific Integrated Circuit \cite{Tofino}. Due to limited hardware resources, we could not fully implement it. However, hardware vendors are currently launching more and more powerful P4-programmable switches, so we are pretty confident that in the future it will be possible to execute P4DDoS in hardware targets, while ensuring a line-rate packet processing speed with only a few-hundreds of nanoseconds packet processing latency, as already possible for simpler strategies \cite{ding2021ivest}.

\vspace{-0.5em}

\section{P4LogLog and P4NEntropy evaluation} \label{sec:evaluation1}
We implemented P4LogLog and P4NEntropy in Python and simulated them for evaluation. 
\subsection{Evaluation metrics and simulation settings}
\subsubsection{Testing flow trace and methodology}
\textbf{P4LogLog}: We use 2018-passive CAIDA flow trace \cite{caida}, including 50 seconds of network traffic, and divide it into 50 1-second time intervals (or observation windows). In each considered time interval there are around 460K packets.\\
\textbf{P4NEntropy}:  We use the same CAIDA flow trace but divide it into 10 observation windows of roughly 5 seconds each, every one including a fixed number of $2^{21}$ packets. Fixing the number of packets per observation window is needed to compare our approach with a state-of-the-art solution \cite{lapolli2019offloading}, named \emph{SOTA\_entropy} for the remainder of the section, which adopts M+A  tables to store pre-computed values and only works with windows with power-of-two number of packets.
 
\subsubsection{Evaluated metrics}
We consider \emph{relative error} as an evaluation metric.\\
\textbf{P4LogLog}: Being $n$ the exact number of distinct flows (either identified by \emph{source IP} or \emph{destination IP} as flow key) in a time interval and $\hat{n}$ its estimated value,  the relative error is defined as  the average value of $\frac{\left| n - \hat{n} \right| }{n} \cdot 100\%$ in all the consecutive 50 time intervals.\\
\textbf{P4NEntropy}: We call $\hat{H}$ the estimated traffic entropy of \emph{destination IPs} in an observation window and $H$ its exact value. The relative error is defined the average value of $\frac{\left| H - \hat{H} \right| }{H} \cdot 100\%$ in the 10 consecutive observation windows. Note that we evaluate the entropy $H$ and not its normalized value $H_{norm}$: this is needed to make a fair comparison with SOTA\_entropy, which does not consider any entropy normalization. To understand how normalization affects the accuracy of the estimated entropy, the reader should refer to the evaluation of P4LogLog (Section \ref{sec:P4LogLog_eval}).  

\subsubsection{Tuning parameters}
The default tuning parameters for P4Log and P4Exp, adopted for both P4LogLog and P4NEntropy, are set as in Tab. \ref{tuning}. The sketch (either Count-min or Count Sketch) used by P4NEntropy has default size ($N_{h}=5$) $\times$ ($N_{s}=2000$). 




\subsection{Evaluation of P4LogLog} \label{sec:P4LogLog_eval}

\begin{figure}[t]
\centering
\subfigure[Flow key: source IP]{
{\scalebox{0.46}{
 \begin{tikzpicture}

\begin{axis}[
legend cell align={left},
legend columns=1,
legend style={font=\large,at={(1, 1)}},
tick align=outside,
tick pos=left,
x grid style={white!69.01960784313725!black},
xlabel={Memory size (Bytes)},
xmajorgrids,
xmin=-0.545, xmax=8.145,
xtick style={color=black},
xtick={0.3,1.3,2.3,3.3,4.3,5.3,6.3,7.3},
xticklabels={10,20,40,80,160,320,640,1280},
y grid style={white!69.01960784313725!black},
ylabel={Relative error (\%)},
ymajorgrids,
ymin=0, ymax=104.9895,
ytick={ 3,20,40,60,80, 100},
ytick style={color=black},
tick label style={font=\large}
]

\draw[fill=blue,draw opacity=0] (axis cs:0.15,0) rectangle (axis cs:0.45,98.72);
\addlegendimage{ybar,ybar legend,fill=blue,draw opacity=0};
\addlegendentry{Linear counting}

\draw[fill=blue,draw opacity=0] (axis cs:1.15,0) rectangle (axis cs:1.45,97.04);
\draw[fill=blue,draw opacity=0] (axis cs:2.15,0) rectangle (axis cs:2.45,93.28);
\draw[fill=blue,draw opacity=0] (axis cs:3.15,0) rectangle (axis cs:3.45,84.94);
\draw[fill=blue,draw opacity=0] (axis cs:4.15,0) rectangle (axis cs:4.45,66.64);
\draw[fill=blue,draw opacity=0] (axis cs:5.15,0) rectangle (axis cs:5.45,26.82);
\draw[fill=blue,draw opacity=0] (axis cs:6.15,0) rectangle (axis cs:6.45,2.55);
\draw[fill=blue,draw opacity=0] (axis cs:7.15,0) rectangle (axis cs:7.45,1.21);
\draw[fill=red,draw opacity=0] (axis cs:0.45,0) rectangle (axis cs:0.75,15.70);
\addlegendimage{ybar,ybar legend,fill=red,draw opacity=0};
\addlegendentry{P4LogLog}

\draw[fill=red,draw opacity=0] (axis cs:1.45,0) rectangle (axis cs:1.75,15.81);
\draw[fill=red,draw opacity=0] (axis cs:2.45,0) rectangle (axis cs:2.75,19.18);
\draw[fill=red,draw opacity=0] (axis cs:3.45,0) rectangle (axis cs:3.75,5.28);
\draw[fill=red,draw opacity=0] (axis cs:4.45,0) rectangle (axis cs:4.75,4.14);
\draw[fill=red,draw opacity=0] (axis cs:5.45,0) rectangle (axis cs:5.75,5.82);
\draw[fill=red,draw opacity=0] (axis cs:6.45,0) rectangle (axis cs:6.75,3.23);
\draw[fill=red,draw opacity=0] (axis cs:7.45,0) rectangle (axis cs:7.75,1.55);
\addplot [semithick, black, dashed, forget plot]
table {%
-0.545 3
8.145 3
};
\end{axis}
\end{tikzpicture}
    \label{fig:sub:car_src}
    }
    }
    }
\subfigure[Flow key: destination IP]{
{\scalebox{0.46}{
\begin{tikzpicture}

\begin{axis}[
legend cell align={left},
legend columns=1,
legend style={font=\large,at={(1, 1)}},
tick align=outside,
tick pos=left,
x grid style={white!69.01960784313725!black},
xlabel={Memory size (Bytes)},
xmajorgrids,
xmin=-0.545, xmax=8.145,
xtick style={color=black},
xtick={0.3,1.3,2.3,3.3,4.3,5.3,6.3,7.3},
xticklabels={10,20,40,80,160,320,640,1280},
y grid style={white!69.01960784313725!black},
ylabel={Relative error (\%)},
ymajorgrids,
ymin=0, ymax=104.9895,
ytick style={color=black},
ytick={3,20,40,60,80, 100},
tick label style={font=\large}
]

\draw[fill=blue,draw opacity=0] (axis cs:0.15,0) rectangle (axis cs:0.45,98.43);
\addlegendimage{ybar,ybar legend,fill=blue,draw opacity=0};
\addlegendentry{Linear counting}

\draw[fill=blue,draw opacity=0] (axis cs:1.15,0) rectangle (axis cs:1.45,96.37);
\draw[fill=blue,draw opacity=0] (axis cs:2.15,0) rectangle (axis cs:2.45,91.76);
\draw[fill=blue,draw opacity=0] (axis cs:3.15,0) rectangle (axis cs:3.45,81.54);
\draw[fill=blue,draw opacity=0] (axis cs:4.15,0) rectangle (axis cs:4.45,59.11);
\draw[fill=blue,draw opacity=0] (axis cs:5.15,0) rectangle (axis cs:5.45,11.34);
\draw[fill=blue,draw opacity=0] (axis cs:6.15,0) rectangle (axis cs:6.45,2.49);
\draw[fill=blue,draw opacity=0] (axis cs:7.15,0) rectangle (axis cs:7.45,1.57);
\draw[fill=red,draw opacity=0] (axis cs:0.45,0) rectangle (axis cs:0.75,17.24);
\addlegendimage{ybar,ybar legend,fill=red,draw opacity=0};
\addlegendentry{P4LogLog}

\draw[fill=red,draw opacity=0] (axis cs:1.45,0) rectangle (axis cs:1.75,13.23);
\draw[fill=red,draw opacity=0] (axis cs:2.45,0) rectangle (axis cs:2.75,10.89);
\draw[fill=red,draw opacity=0] (axis cs:3.45,0) rectangle (axis cs:3.75,8.21);
\draw[fill=red,draw opacity=0] (axis cs:4.45,0) rectangle (axis cs:4.75,5.41);
\draw[fill=red,draw opacity=0] (axis cs:5.45,0) rectangle (axis cs:5.75,3.64);
\draw[fill=red,draw opacity=0] (axis cs:6.45,0) rectangle (axis cs:6.75,3.54);
\draw[fill=red,draw opacity=0] (axis cs:7.45,0) rectangle (axis cs:7.75,2.26);
\addplot [semithick, black, dashed, forget plot]
table {%
-0.545 3
8.145 3
};
\end{axis}

\end{tikzpicture}
    \label{fig:sub:car_dst}
    }
    }
    }
\vspace{-1em}
\caption{Performance comparison of P4LogLog with an existing flow cardinality estimation approach \cite{yu2013software}}
\label{carcomparision}
\end{figure}
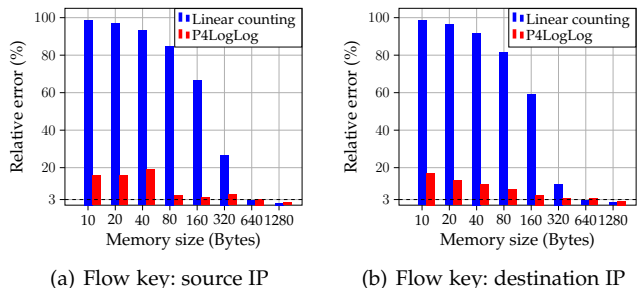

As shown in Fig. \ref{carcomparision}, we compare our P4LogLog with another existing flow cardinality estimator (\emph{Linear counting} \cite{yu2013software}), implementable in a programmable data plane, in terms of relative error.
1 bit is used for each Linear counting register cell \cite{whang1990linear}, while 5 bits are allocated for each P4LogLog register cell \cite{durand2003loglog}. Given this, we vary the \emph{memory size} of each register for both the approaches (i.e., we vary the number of cells in the registers, which can be easily retrieved).

Figure \ref{fig:sub:car_src} focuses on the estimation of distinct source IPs in the trace. The relative error on such a flow cardinality estimation by adopting Linear counting is 50\% higher than by adopting P4LogLog when the memory size is below 320 bytes, and its value for Linear counting is high for any memory size below 640 bytes. Conversely, our P4LogLog leads to acceptable relative errors with only 80 bytes. If we assign 1280-bytes registers to P4LogLog and Linear counting, the relative error of both is around 1\%. 
Likewise, Fig. \ref{fig:sub:car_dst} shows the estimated number of distinct destination IPs in the trace. Our P4LogLog algorithm still outperforms Linear counting for small memory sizes. 
When the memory occupation reaches 640 bytes, the relative error of P4LogLog is below 3\%, which is assumed as an acceptable target. 

Another solution for flow cardinality estimation is proposed in \cite{liu2016one}. However, such a solution always needs much more memory than Linear Counting and P4LogLog (i.e., at least 0.2MB)  to get reasonable accuracy.

\subsection{Evaluation of P4NEntropy}
We simulate both our strategy and SOTA\_entropy in the case that the packet count of each flow (identified by destination IP flow key) is estimated in the data plane by adopting either Count-min Sketch or Count Sketch (see Section \ref{sec:background_sketch} and Fig. \ref{nwh}). 

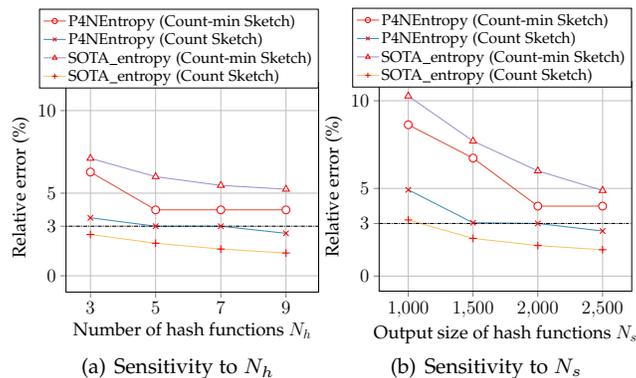
\begin{figure}[t]
\centering
\subfigure[Sensitivity to $N_{h}$]{
\hspace{-1em}
{\scalebox{0.49}{
\begin{tikzpicture}

\definecolor{color0}{rgb}{0.886274509803922,0.290196078431373,0.2}
\definecolor{color1}{rgb}{0.203921568627451,0.541176470588235,0.741176470588235}
\definecolor{color2}{rgb}{0.596078431372549,0.556862745098039,0.835294117647059}
\definecolor{color3}{rgb}{0.984313725490196,0.756862745098039,0.368627450980392}

\begin{axis}[
legend cell align={left},
legend entries={{P4NEntropy (Count-min Sketch)},{P4NEntropy (Count Sketch)},{SOTA\_entropy (Count-min Sketch)}, {SOTA\_entropy (Count Sketch)}},
legend style={font=\large, at={(1, 1.35)}},
tick align=outside,
tick pos=left,
xlabel={Number of hash functions $N_{h}$},
xmajorgrids,
xmin=2.25, xmax=10,
xtick={0,3,5,7,9},
ylabel={Relative error (\%)},
ymajorgrids,
ytick={0,3,5,10},
ymin=-0.986, ymax=11.706
]
\addlegendimage{mark=o,mark options={color=red}, color0}
\addlegendimage{mark=x,mark options={color=red}, color1}
\addlegendimage{mark=triangle,mark options={color=red}, color2}
\addlegendimage{mark=+,mark options={color=red}, color3}

\addplot [semithick, color0, mark=*, mark size=3, mark options={solid,fill=white,draw=red}]
table [row sep=\\]{%
3	6.28 \\
5	3.99 \\
7	3.99 \\
9	3.99 \\
};
\addplot [semithick, color1, mark=x, mark size=3, mark options={solid,fill=white,draw=red}]
table [row sep=\\]{%
3	3.51 \\
5	3.00 \\
7	3.00 \\
9	2.57 \\
};
\addplot [semithick, color2, mark=triangle*, mark size=3, mark options={solid,fill=white,draw=red}]
table [row sep=\\]{%
3	7.11 \\
5	6.00 \\
7	5.47 \\
9	5.24 \\
};

\addplot [semithick, color3, mark=+, mark size=3, mark options={solid,fill=white,draw=red}]
table [row sep=\\]{%
3	2.50 \\
5	1.97 \\
7	1.62 \\
9	1.38 \\
};
\addplot [semithick, black,  densely dashdotted, forget plot]
table [row sep=\\]{%
1	3 \\
26	3 \\
};
\end{axis}

\end{tikzpicture}
    \label{fig:sub:Nh}
    }
    }
    }
\hspace{-2em}
\subfigure[Sensitivity to $N_{s}$]{
{\raisebox{-0.2em}{
\scalebox{0.49}{
\begin{tikzpicture}

\definecolor{color0}{rgb}{0.886274509803922,0.290196078431373,0.2}
\definecolor{color1}{rgb}{0.203921568627451,0.541176470588235,0.741176470588235}
\definecolor{color2}{rgb}{0.596078431372549,0.556862745098039,0.835294117647059}
\definecolor{color3}{rgb}{0.984313725490196,0.756862745098039,0.368627450980392}

\begin{axis}[
legend cell align={left},
legend entries={{P4NEntropy (Count-min Sketch)},{P4NEntropy (Count Sketch)},{SOTA\_entropy (Count-min Sketch)}, {SOTA\_entropy (Count Sketch)}},
legend style={font=\large, at={(1, 1.35)}},
tick align=outside,
tick pos=left,
xlabel={{Output size of hash functions $N_{s}$}},
xmajorgrids,
xmin=775, xmax=2725,
ylabel={{Relative error (\%)}},
ytick={0,3,5,10},
ymajorgrids,
ymin=-0.861, ymax=11.101
]
\addlegendimage{mark=o,mark options={color=red}, color0}
\addlegendimage{mark=x,mark options={color=red}, color1}
\addlegendimage{mark=triangle,mark options={color=red}, color2}
\addlegendimage{mark=+,mark options={color=red}, color3}

\addplot [semithick, color0, mark=*, mark size=3, mark options={solid,fill=white,draw=red}]
table [row sep=\\]{%
1000	    8.63 \\
1500	6.73 \\
2000	3.99 \\
2500	3.99 \\
};
\addplot [semithick, color1, mark=x, mark size=3, mark options={solid,fill=white,draw=red}]
table [row sep=\\]{%
1000 	4.92 \\
1500	3.04 \\
2000	3.00 \\
2500	2.57 \\
};

\addplot [semithick, color2, mark=triangle*, mark size=3, mark options={solid,fill=white,draw=red}]
table [row sep=\\]{%
1000	10.27 \\
1500	7.70 \\
2000	6.00 \\
2500	4.88 \\
};
\addplot [semithick, color3, mark=+, mark size=3, mark options={solid,fill=white,draw=red}]
table [row sep=\\]{%
1000	    3.20 \\
1500	2.15 \\
2000	1.74 \\
2500	1.50 \\
};
\addplot [semithick, black,  densely dashdotted, forget plot]
table [row sep=\\]{%
776	3 \\
2700	3 \\
};
\end{axis}

\end{tikzpicture}
    \label{fig:sub:Ns}
    }
    }
    }
    }
\vspace{-1em}
\caption{Performance comparison of P4NEntropy with an existing entropy estimation approach \protect\cite{lapolli2019offloading} (flow key: destination IP)}
\label{entropyComparison}
\end{figure}

We show how entropy estimation of destination IPs is affected while changing the size $N_h\times N_s$ of the sketch (Fig. \ref{entropyComparison}). Fig. \ref{fig:sub:Nh} shows the relative error in entropy estimation for the two strategies when $N_s$ is fixed and $N_h$ varies. It shows that the relative error slightly decreases as $N_{h}$ increases in all the cases. Moreover, P4NEntropy and SOTA\_entropy lead to similar relative error. 
It can be noted that, when adopting Count-min Sketch, both P4NEntropy and SOTA\_entropy  have larger relative error (always above 4\%) than when adopting Count-Sketch. Additionally, in this case, the relative error of SOTA\_entropy is slightly higher than the one of P4NEntropy, which is caused by the different ways how $Sum(f_{i})$ is estimated. In SOTA\_entropy, the Longest Prefix Match (LPM) lookup table for $F(f_{i}) = f_{i} \log_{2}f_{i} - (f_{i} - 1)\log_2(f_{i} - 1)$ (see \cite{lapolli2019offloading}) is sensitive to the large packet count ($f_{i}$) overestimation caused by Count-min Sketch. Conversely, P4NEntropy needs to calculate  $\log_{2}f_i + \frac{1}{\ln2}$ (see Eq. \ref{eq1}), which is less sensitive to large overestimations \emph{(i)} due to the logarithm nature and \emph{(ii)} because $\frac{1}{\ln2}$ is a constant value. This effect does not happen when Count-Sketch is adopted, since overestimations are much less frequent. In that case, P4NEntropy leads to slightly worse results than SOTA\_entropy because, unlike SOTA\_entropy, it uses an approximation for the computation of entropy (see Eq. \ref{eq2}).

Fig. \ref{fig:sub:Ns} shows instead the impact of a variation of $N_s$ on relative error in entropy estimation. Results are similar to what shown in  Fig. \ref{fig:sub:Nh}, but it can be noted that both strategies are more sensitive to a variation of $N_s$ than of $N_h$. In this case, when adopting Count Sketch, relative error is always close to 3\%.
Note that a relative error of 3\% is the maximum possible value ensuring that accuracy of practical monitoring applications is not affected \cite{lall2006data}.

\vspace{-0.5em}
\section{P4DDoS evaluation} \label{sec:evaluation2}
We implemented P4DDoS in Python and simulated it for evaluation. Additionally, we also implemented a state-of-the-art entropy-based DDoS detection approach \cite{lapolli2019offloading} executable in programmable switches, named \emph{SOTA\_DDoS} for the sake of brevity, and compared them. 
To make a fair comparison, both DDoS detection strategies have been implemented leveraging our proposed P4NEntropy strategy and using a sketch, for packet count estimation, of the same size. Note, however, that the original version of SOTA\_DDoS uses SOTA\_Entropy for entropy estimation (see the previous subsection).
Unlike P4DDoS, which triggers a DDoS alarm only when the \emph{normalized} entropy of destination IPs decreses below a threshold, SOTA\_DDoS triggers a DDoS alarm when any of two conditions holds: \emph{(i)}  entropy (not normalized) of source IPs increases above an adaptive threshold and \emph{(ii)} entropy (not normalized) of destination IPs decreases below an adaptive threshold.

\subsection{Evaluation metrics and simulation settings}
\subsubsection{Testing flow trace and methodology}
We consider three kinds of flow traces.\\
\textbf{\emph{Trace1}. Legitimate flow trace:} The same CAIDA flow trace \cite{caida} that we used for the evaluation of P4LogLog. The 50-seconds flow trace is divided into 50 1-second time intervals.\\
\textbf{\emph{Trace2}. Legitimate flow trace mixed with Booter DDoS attack traffic \cite{santannajjIM2015}:}  50-seconds traces taken from a set of Booter DDoS attack traces, and split into 50 time intervals. Each 1-second attack trace is injected into the legitimate 50-seconds flow trace according to its sequential 1-second time intervals. We took four different packet-rate Booter DDoS attack traces into consideration: Tab. \ref{table:flow} reports their properties and names as specified in \cite{santannajjIM2015}. We considered the four traces with the highest number of attack source IPs: this allows us to analyze DDoS attacks with different volumes. Moreover, we also injected all four Booter DDoS attack traces together into the legitimate flow trace: 
we name this trace as \emph{Mixed}. Such a mixed DDoS attack flow trace can help us evaluate the performance of DDoS detection when multiple DDoS attacks occur simultaneously in the network. \\
\textbf{\emph{Trace3}. Legitimate flow trace mixed with internal Botnet DDoS attack traffic:} In this case, we assume that some internal hosts of the network (e.g., a datacenter network) 
are exploited by an attacker to reverse malicious traffic towards a DDoS victim within the same network. We varied the \emph{attack traffic proportion} (i.e., the percentage of generated malicious traffic over the total traffic in the network) from 5\% to 30\%. This flow trace is generated by crafting Trace 1 in such a way that part of the traffic is forwarded to one specific DDoS victim (by changing the destination IP of a given proportion of the packets).

\begin{table}[t]
\caption{Properties of DDoS flow traces \cite{santannajjIM2015}}
\label{table:flow}
\centering
\vspace{-1em}
\begin{adjustbox}{width=0.45\textwidth}
\begin{tabular}{|c|c|c|}
\hline
\textbf{DDoS trace name} & \textbf{Packet per second} & \textbf{Attack source IPs}\\
\hline
Booter 6 & $\sim$ 90000 & 7379\\
\hline
Booter 7  & $\sim$ 41000 & 6075\\
\hline
Booter 1 & $\sim$ 96000 & 4486\\
\hline
Booter 4 & $\sim$ 80000 & 2970\\
\hline
\end{tabular}
\end{adjustbox}
\end{table}	

\subsubsection{Evaluation metrics}

We consider \emph{true-positive rate} $D_{tp}$, \emph{false-positive rate} $D_{fp}$ and \emph{detection accuracy} $D_{acc}$ as evaluation metrics. Considering that \emph{(i)} True Positive (TP) is the number of time intervals with a triggered DDoS alarm while a DDoS attack is occurring in those intervals,  \emph{(ii)} True Negative (TN) is the number of time intervals without any triggered DDoS alarm while no DDoS attack is occurring,  \emph{(iii)} False Positive (FP) is the number of time intervals with a triggered DDoS alarm while no DDoS attack is occurring, and \emph{(iv)} False Negative (FN) is the number of time intervals without any triggered DDoS alarm while a DDoS attack is instead occurring, the metrics introduced above are defined as:

{\small
\begin{align*}
    D_{tp} &= \frac{TP}{TP+FN} \times 100\%\\
   D_{fp} &= \frac{FP }{TN+FP} \times 100\%\\
    D_{acc} &= \frac{TP+TN}{TP+TN+FP+FN} \times 100\%
\end{align*}}

\subsubsection{Tuning parameters}
The smoothing factor in $EWMA_{norm}$ and for the thresholds defined in SOTA\_DDoS is set to $\alpha=0.13$: with this value, all the previous computed averages (up to all the 50 time intervals) have some impact on EWMA. 
All the parameters for P4Log and P4Exp are the ones reported in Tab. \ref{tuning}. We choose Count Sketch as sketch for P4NEntropy, with $(N_{h}=5) \times (N_{s}=2000)$. The register size in P4LogLog is set to $m=2048$, which corresponds to 1280 Bytes of memory. The considered time intervals $T_{int}$, as already said, are 1-second wide. With longer $T_{int}$, $N_h$ and $N_s$ should be properly increased to ensure good entropy estimation accuracy. Finally, the normalized entropy parameter is set to $\epsilon=0.01$ unless otherwise specified. 

\begin{table*}[t]
 \caption{Comparison of P4DDoS detection performance with a state-of-the-art approach \cite{lapolli2019offloading} (Booter DDoS attacks)}
\label{table2}
 \centering
 \vspace{-1em}
\begin{adjustbox}{width=0.67\textwidth}
\begin{tabular}{|c|c|c|c|c|c|c|}
\hline
\multirow{2}{*}{\textbf{Algorithm} } & \multirow{2}{*}{\begin{tabular}[x]{@{\vspace{-0.1cm}}c@{}}\textbf{False-positive }\\\textbf{rate} $D_{fp}$\end{tabular} } & \multicolumn{5}{c|}{\textbf{True-positive rate} $D_{tp}$ / \textbf{Detection accuracy} $D_{acc}$} \\
\cline{3-7}
~ & ~ &\textbf{Booter 6} & \textbf{Booter 7} & \textbf{Booter 1} & \textbf{Booter 4} & \textbf{Mixed} \\
\hline
P4DDoS & 8\% & 100\% / 96\% & 82\% / 87\%& 96\% / 94\% & 98\% / 95\%& 100\% / 96\% \\
\hline
\begin{tabular}[x]{@{\vspace{-0.1cm}}c@{}}SOTA\_DDoS\\(k=5.5)\end{tabular}& 6\% & 96\% / 95\% & 32\% / 63\% & 62\% / 78\% & 70\% / 82\%& 100\% / 97\% \\ [1.3ex]
\hline
\begin{tabular}[x]{@{\vspace{-0.1cm}}c@{}}SOTA\_DDoS\\(k=4.5)\end{tabular}& 8\% & 100\% / 96\% & 38\% / 65\% & 82\% / 87\% & 78\% / 85\%& 100\% / 96\% \\ [1.3ex]
\hline
\begin{tabular}[x]{@{\vspace{-0.1cm}}c@{}}SOTA\_DDoS\\(k=3.5)\end{tabular}& 10\% & 100\% / 95\% & 74\% / 82\% & 100\% / 95\% & 94\% / 92\%& 100\% / 95\% \\ [1.3ex]
\hline
\begin{tabular}[x]{@{\vspace{-0.1cm}}c@{}}SOTA\_DDoS\\(k=2.5)\end{tabular}& 20\% & 100\% / 90\% & 94\% / 87\%& 100\% / 90\% &100\% / 90\%& 100\% / 90\%\\ [1.3ex]
\hline
\begin{tabular}[x]{@{\vspace{-0.1cm}}c@{}}SOTA\_DDoS\\(k=1.5)\end{tabular}& 38\% & 100\% / 81\% & 100\% / 81\%& 100\% / 81\% & 100\% / 81\%& 100\% / 81\% \\ [1.3ex]
\hline
\begin{tabular}[x]{@{\vspace{-0.1cm}}c@{}}SOTA\_DDoS\\(k=0.5)\end{tabular} & 60\% & 100\% / 70\% & 100\% / 70\%& 100\% / 70\% & 100\% / 70\%& 100\% / 70\% \\ [1.3ex]
\hline
\end{tabular}
\end{adjustbox}
\end{table*}

\begin{figure*}[t]
\centering
\subfigure[Sensitivity to false-positive rate]{
{\scalebox{0.5}{
\begin{tikzpicture}

\definecolor{color0}{rgb}{0.12156862745098,0.466666666666667,0.705882352941177}
\definecolor{color1}{rgb}{1,0.498039215686275,0.0549019607843137}
\definecolor{color2}{rgb}{0.172549019607843,0.627450980392157,0.172549019607843}
\definecolor{color3}{rgb}{0.83921568627451,0.152941176470588,0.156862745098039}
\definecolor{color4}{rgb}{0.580392156862745,0.403921568627451,0.741176470588235}

\begin{axis}[
legend cell align={left},
legend columns=3,
legend style={fill opacity=0.8, draw opacity=1, text opacity=1, at={(0.05,1)}, anchor=south west, draw=white!80.0!black},
tick align=outside,
tick pos=left,
x grid style={white!69.01960784313725!black},
xlabel={  $\epsilon$},
xmajorgrids,
xmin=-0.0055, xmax=0.1045,
xtick style={color=black},
y grid style={white!69.01960784313725!black},
ylabel={False postive rate (\%)},
ymajorgrids,
ytick style={color=black},
ymin=-5, ymax=105,
ytick={0, 8, 20, 32, 40, 60, 80, 100},
scaled ticks=false,
tick label style={/pgf/number format/fixed} 
]
\addplot [semithick, color0, mark=*, mark size=3, mark options={solid,fill=white,draw=red}]
table {%
0 32
0.01 8
0.02 8
0.03 2
0.04 0
0.05 0
0.06 0
0.07 0
0.08 0
0.09 0
0.1 0
};
\end{axis}

\end{tikzpicture}
    \label{fig:sub:senTheta}
    }
    }
    }
\subfigure[Sensitivity to true-positive rate]{
{\scalebox{0.5}{
\begin{tikzpicture}

\definecolor{color0}{rgb}{0.12156862745098,0.466666666666667,0.705882352941177}
\definecolor{color1}{rgb}{1,0.498039215686275,0.0549019607843137}
\definecolor{color2}{rgb}{0.172549019607843,0.627450980392157,0.172549019607843}
\definecolor{color3}{rgb}{0.83921568627451,0.152941176470588,0.156862745098039}
\definecolor{color4}{rgb}{0.580392156862745,0.403921568627451,0.741176470588235}

\begin{axis}[
legend cell align={left},
legend columns=3,
legend style={font=\Large,fill opacity=0.8, draw opacity=1, text opacity=1, at={(-0.15, 1)},
anchor=south west, draw=white!80.0!black},
tick align=outside,
tick pos=left,
x grid style={white!69.01960784313725!black},
xlabel={$\epsilon$},
xmajorgrids,
xmin=-0.0055, xmax=0.1045,
xtick style={color=black},
y grid style={white!69.01960784313725!black},
ylabel={True positive rate (\%)},
ymajorgrids,
ymin=-5, ymax=105,
ytick style={color=black},
scaled ticks=false,
tick label style={/pgf/number format/fixed} 
]
\addplot [semithick, color0, mark=*, mark size=3, mark options={solid,fill=white,draw=red}]
table {%
0 100
0.01 100
0.02 98
0.03 80
0.04 80
0.05 80
0.06 78
0.07 42
0.08 42
0.09 36
0.1 8
};
\addlegendentry{Booter 6}
\addplot [semithick, color1, mark=x, mark size=3, mark options={solid,fill=white,draw=red}]
table {%
0 82
0.01 82
0.02 80
0.03 18
0.04 18
0.05 18
0.06 6
0.07 0
0.08 0
0.09 0
0.1 0
};
\addlegendentry{Booter 7}
\addplot [semithick, color2, mark=triangle*, mark size=3, mark options={solid,fill=white,draw=red}]
table {%
0 96
0.01 96
0.02 96
0.03 68
0.04 68
0.05 68
0.06 58
0.07 30
0.08 30
0.09 26
0.1 4
};
\addlegendentry{Booter 1}
\addplot [semithick, color3, mark=asterisk, mark size=3, mark options={solid,fill=white,draw=red}]
table {%
0 98
0.01 98
0.02 96
0.03 42
0.04 42
0.05 42
0.06 32
0.07 8
0.08 8
0.09 6
0.1 2
};
\addlegendentry{Booter 4}
\addplot [semithick, color4, mark=+, mark size=3, mark options={solid,fill=white,draw=red}]
table {%
0 100
0.01 100
0.02 100
0.03 100
0.04 100
0.05 100
0.06 100
0.07 100
0.08 100
0.09 100
0.1 100
};
\addlegendentry{$Mixed$}
\end{axis}

\end{tikzpicture}
    \label{fig:sub:sensEp}
    }
    }
    }
\subfigure[Sensitivity to detection accuracy]{
{\scalebox{0.5}{
\begin{tikzpicture}

\definecolor{color0}{rgb}{0.12156862745098,0.466666666666667,0.705882352941177}
\definecolor{color1}{rgb}{1,0.498039215686275,0.0549019607843137}
\definecolor{color2}{rgb}{0.172549019607843,0.627450980392157,0.172549019607843}
\definecolor{color3}{rgb}{0.83921568627451,0.152941176470588,0.156862745098039}
\definecolor{color4}{rgb}{0.580392156862745,0.403921568627451,0.741176470588235}

\begin{axis}[
legend cell align={left},
legend columns=3,
legend style={font=\Large, fill opacity=0.8, draw opacity=1, text opacity=1, at={(-0.15, 1)},
anchor=south west, draw=white!80.0!black},
tick align=outside,
tick pos=left,
x grid style={white!69.01960784313725!black},
xlabel={$\epsilon$},
xmajorgrids,
xmin=-0.0055, xmax=0.1045,
xtick style={color=black},
y grid style={white!69.01960784313725!black},
ylabel={Detection accuracy (\%)},
ymajorgrids,
ymin=-5, ymax=105,
ytick style={color=black},
scaled ticks=false,
tick label style={/pgf/number format/fixed} 
]
\addplot [semithick, color0, mark=*, mark size=3, mark options={solid,fill=white,draw=red}]
table {%
0 84
0.01 96
0.02 95
0.03 89
0.04 90
0.05 90
0.06 89
0.07 71
0.08 71
0.09 68
0.1 54
};
\addlegendentry{Booter 6}
\addplot [semithick, color1, mark=x, mark size=3, mark options={solid,fill=white,draw=red}]
table {%
0 75
0.01 87
0.02 86
0.03 58
0.04 59
0.05 59
0.06 53
0.07 50
0.08 50
0.09 50
0.1 50
};
\addlegendentry{Booter 7}
\addplot [semithick, color2, mark=triangle*, mark size=3, mark options={solid,fill=white,draw=red}]
table {%
0 82
0.01 94
0.02 94
0.03 83
0.04 84
0.05 84
0.06 79
0.07 65
0.08 65
0.09 63
0.1 52
};
\addlegendentry{Booter 1}
\addplot [semithick, color3, mark=asterisk, mark size=3, mark options={solid,fill=white,draw=red}]
table {%
0 83
0.01 95
0.02 94
0.03 70
0.04 71
0.05 71
0.06 66
0.07 54
0.08 54
0.09 53
0.1 51
};
\addlegendentry{Booter 4}
\addplot [semithick, color4, mark=+, mark size=3, mark options={solid,fill=white,draw=red}]
table {%
0 84
0.01 96
0.02 96
0.03 98
0.04 100
0.05 100
0.06 100
0.07 100
0.08 100
0.09 100
0.1 100
};
\addlegendentry{$Mixed$}
\end{axis}

\end{tikzpicture}
    \label{fig:sub:sensAcc}
    }
    }
    }
\caption{Sensitivity analysis of P4DDoS to parameter $\epsilon$}
\label{sensTandE}
\end{figure*}
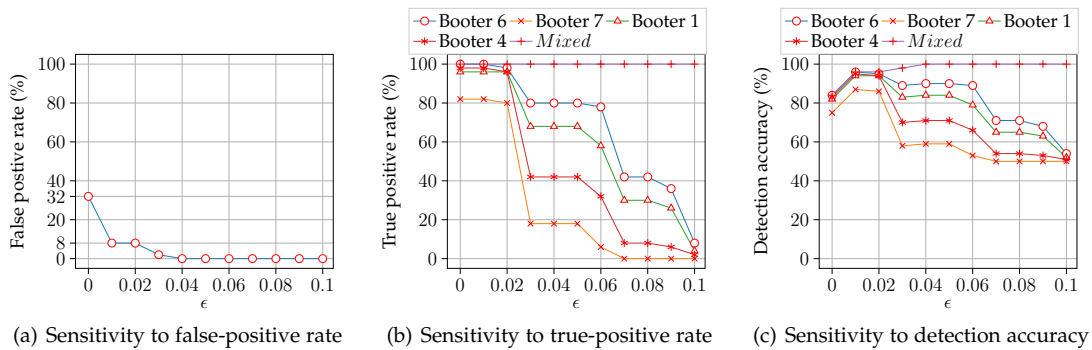

\subsection{Detection performance (Booter DDoS attacks)} \label{sec:booter}
In this subsection, we evaluate our P4DDoS strategy against the state-of-the-art approach SOTA\_DDoS in terms of $D_{tp}$, $D_{fp}$ and $D_{acc}$ in the case of Booter DDoS attacks. We also perform a sensitivity analysis of P4DDoS against the parameter $\epsilon$, showing how the detection performance is affected by changing its value. The testing flow trace is composed by the concatenation of \emph{Trace1} and \emph{Trace2}: we first run 50-seconds legitimate flow trace (\emph{Trace 1}) so that adaptive thresholds on entropy, for both strategies, are properly set in a legitimate traffic scenario. This trace allows us to evaluate $D_{fp}$.
Then, \emph{Trace 2} including different packet-rate DDoS attacks (also mixed), is used to evaluate $D_{tp}$ and, together with results obtained in \emph{Trace 1},  $D_{acc}$. 

\subsubsection{Comparison with the state of the art} 
\begin{table*}[t]
 \caption{Comparison of P4DDoS detection performance with a state-of-the-art approach \cite{lapolli2019offloading} for different Botnet DDoS attack traffic proportions (ATPs)}
\label{table3}
 \centering
 \vspace{-1em}
\begin{adjustbox}{width=0.75\textwidth}
\begin{tabular}{|c|c|c|c|c|c|c|c|}
\hline
\multirow{2}{*}{\textbf{Algorithm} } & \multirow{2}{*}{\begin{tabular}[x]{@{\vspace{-0.1cm}}c@{}}\textbf{False positive }\\\textbf{rate} $D_{fp}$\end{tabular} } & \multicolumn{6}{c|}{\textbf{True-positive rate} $D_{tp}$ / \textbf{Detection accuracy} $D_{acc}$} \\
\cline{3-8}
~ & ~ &\textbf{ATP: 5\%} & \textbf{ATP: 10\%} & \textbf{ATP: 15\%} & \textbf{ATP: 20\%} & \textbf{ATP: 25\%} & \textbf{ATP: 30\%}\\
\hline
P4DDoS & 8\% & 36\% / 64\% & 92\% / 92\%& 100\% / 96\% & 100\% / 96\%& 100\% / 96\%&100\% / 96\% \\
\hline
\begin{tabular}[x]{@{\vspace{-0.1cm}}c@{}}SOTA\_DDoS\\(k=5.5)\end{tabular}& 6\% & 0\% / 47\% & 12\% / 53\% & 68\% / 81\% & 100\% / 97\%& 100\% / 97\% &100\% / 97\%\\ [1.3ex]
\hline
\begin{tabular}[x]{@{\vspace{-0.1cm}}c@{}}SOTA\_DDoS\\(k=4.5)\end{tabular}& 8\% & 0\% / 46\% & 40\% /  66\% & 96\% / 94\% & 100\% / 96\%& 100\% / 96\% & 100\% / 96\%\\ [1.3ex]
\hline
\begin{tabular}[x]{@{\vspace{-0.1cm}}c@{}}SOTA\_DDoS\\(k=3.5)\end{tabular}& 10\% & 10\% / 50\% & 50\% / 70\% & 100\% /  95\% & 100\% / 95\%& 100\% / 95\% & 100\% / 95\%\\ [1.3ex]
\hline
\begin{tabular}[x]{@{\vspace{-0.1cm}}c@{}}SOTA\_DDoS\\(k=2.5)\end{tabular}& 20\% & 20\% / 50\% & 88\% / 84\%& 100\% / 90\% & 100\% / 90\%& 100\% / 90\%& 100\% / 90\%\\ [1.3ex]
\hline
\begin{tabular}[x]{@{\vspace{-0.1cm}}c@{}}SOTA\_DDoS\\(k=1.5)\end{tabular}& 38\% & 82\% / 72\% & 94\% / 78\%& 100\% / 81\% & 100\% / 81\%& 100\% / 81\%& 100\% / 81\% \\ [1.3ex]
\hline
\begin{tabular}[x]{@{\vspace{-0.1cm}}c@{}}SOTA\_DDoS\\(k=0.5)\end{tabular} & 60\% & 96\% / 68\% & 100\% / 70\%& 100\% / 70\% & 100\% / 70\%& 100\% / 70\%& 100\% / 70\%\\ [1.3ex]
\hline
\end{tabular}
\end{adjustbox}
\end{table*}
To fairly compare P4DDoS with SOTA\_DDoS, we tuned the sensitivity coefficient $k$ of SOTA\_DDoS (see \cite{lapolli2019offloading}) to different values: lower $k$ 
leads to higher true-positive rate but also higher false-positive rate. 
Evaluation results are reported in Tab. \ref{table2}.  In the first 50 time intervals, four false alarms are detected by P4DDoS, being thus the false-positive rate 8\%. As said, the false-positive rate of SOTA\_DDoS increases as $k$ decreases. 
False-positive rate of P4DDoS is slightly higher than of SOTA\_DDoS only when $k=5.5$ but, in that case, P4DDoS outperforms SOTA\_DDoS on both true-positive rate and detection accuracy for all the considered Booter attacks. The best trade-off between all the metrics for SOTA\_DDoS is obtained with $k=3.5$. In this case, P4DDoS and SOTA\_DDoS have comparable performance (with slightly better performance for P4DDoS). This means that, in this scenario, comparing the normalized entropy of destination IPs against a well-defined threshold is enough to get good performance on DDoS detection and that an evaluation of entropy of source IPs can be avoided (that is, same performance can be obtained with a simpler strategy). 
 
 

\subsubsection{Sensitivity analysis} \label{sensDDoS}
Figure \ref{sensTandE} reports the sensitivity of P4DDoS to normalized network traffic entropy parameter $\epsilon$.
Figure \ref{fig:sub:senTheta} shows that false-positive rate 
decreases as $\epsilon$ is smaller and stabilizes to zero once $\epsilon$ is larger than 0.04. This is because larger $\epsilon$ results in a smaller threshold, being more DDoS alarms triggered also when DDoS attacks are not occurring. False positives only happen  for the legitimate traffic, reason why only one curve is reported.
 Figure \ref{fig:sub:sensEp} reveals the behavior of true-positive rate when $\epsilon$ varies, showing that in general true-positive rate decreases as $\epsilon$ increases. 
Figure \ref{fig:sub:sensAcc} shows the impact of $\epsilon$ on detection accuracy. The shown curves, apart from the Mixed case, have a maximum at around $\epsilon=0.01$: we then decided to set $\epsilon$ to this value, since it leads to the best trade-off considering all the three metrics.

\subsection{Detection performance (Botnet DDoS attacks)}
Table \ref{table3} shows a comparison on DDoS detection performance in case of internal Botnet DDoS attacks. The same methodology as described in Section \ref{sec:booter} is adopted to prepare the testing flow trace but, in this case, \emph{Trace1} and \emph{Trace3} are concatenated. 
In this attack scenario, the cardinality of source IPs in the network does not change and
the attack traffic proportion in Trace 3 is varied from 5\% to 30\%.
Intuitively, the detection accuracy of P4DDoS increases as the attack traffic proportion increases. When the attack traffic rate is low, i.e., 5\%, the true-positive rate of P4DDoS is 36\%. This is the drawback of most normalized entropy-based DDoS detection strategies: they struggle to detect low-packet-rate DDoS attacks since the normalized entropy may not significantly decrease. 
Nevertheless, our P4DDoS still has higher (or at least comparable) detection accuracy than SOTA\_DDoS for any coefficient $k$. This is due to the fact that the entropy of destination IPs (not normalized) may decrease because of either a decrease in the cardinality of destination IPs in consecutive time intervals (see Section  \ref{sec:entropy}) or because a DDoS attack is occurring. Instead, the normalized entropy (used by P4DDoS) decreases only when a DDoS attack is occurring, since it is normalized to the cardinality of destination IPs. Thus, by considering non-normalized entropy as the metric to detect DDoS attacks as done by SOTA\_DDoS, there is a higher chance of false positives due to legitimate traffic oscillations in consecutive time intervals. It is also important to note that the entropy of source IPs may not significantly increase when a Botnet DDoS attack occurs (as proven in \cite{bhandari2015destination}), so a simpler entropy-based DDoS detection system only considering normalized entropy of destination IPs may suffice for the detection of a wide range of attacks.
\vspace{-0.5em}
\section{Related work}\label{sec:related}

Here we recall existing works on \emph{flow cardinality estimation} and on \emph{entropy-based DDoS detection} in Software-Defined Networks with programmable data planes. 
 

\textbf{Flow cardinality estimation for network monitoring}: 
Many cardinality-estimation algorithms have been implemented to be executed in programmable data planes for the purpose of network monitoring \cite{yu2013software}\cite{liu2016one}\cite{yang2018elastic}, often based on linear counting \cite{whang1990linear}. However, all of them are able to only perform the update operation directly in the data plane, while the query operation has still to be executed by the controller. This is because programmable switches do not support arithmetic operations such as logarithm and exponential function computation, which are needed for flow cardinality estimation. Conversely, by leveraging our proposed strategies for logarithm and exponential-function estimation in the data plane, named P4Log and P4Exp \cite{ding2019estimating}, we developed P4LogLog, a flow cardinality estimation algorithm that takes inspiration from LogLog \cite{durand2003loglog}. P4LogLog enables a flow cardinality estimation entirely in programmable switches, where both update and query operations can be executed in the data plane. 
Moreover, our P4LogLog can estimate cardinality with high accuracy while consuming less memory than existing approaches.  Note that HyperLogLog \cite{flajolet2007hyperloglog} has higher theoretical accuracy than LogLog, but it is currently not implementable in P4 language due to the computation of harmonic mean. 
HyperLogLog can also be implemented in CPU-based and FPGA-based programmable data planes \cite{kulkarni2020hyperloglog}. However, the achievable throughput is  limited and the adopted language is target-specific, while P4 can be used to program the data plane pipeline of heterogeneous hardware/software targets.


 \textbf{Entropy-based DDoS detection}: 
 Entropy-based DDoS detection has been widely studied in the context of SDN:  a significant decrease in the (normalized) network entropy of destination IPs in a given time interval can be an indication of occurrence of a DDoS attack \cite{giotis2014combining}\cite{kalkan2018jess}\cite{wang2015entropy}\cite{mousavi2015early}. However, in most of previous works, entropy estimation is executed by the controller due to the complex way it is computed. 
Some works can be found in literature dealing with network traffic entropy estimation performed partially in the switches' data plane. For example, papers \cite{liu2016one}\cite{huang2017sketchvisor}\cite{yang2018elastic} all envision some operations to be executed by the programmable data plane, so that only summarized data must be sent to the controller. However, since the controller needs to frequently retrieve information from all the switches, the generated communication overhead is significant. 
 Recently, Lapolli \textit{et al.} \cite{lapolli2019offloading} have demonstrated the feasibility of performing network traffic entropy estimation in the data plane using the P4 language, with the aim of detecting DDoS attacks. Their approach is valuable but it requires the usage of TCAM, which is instead avoided by our proposed P4DDoS. Moreover, P4DDoS and P4NEntropy adopt a time-based observation window, while \cite{lapolli2019offloading} requires an observation window that includes a fixed power-of-two number of packets, making their solution less flexible. In fact, our approach may allow a controller to synchronize the retrieval of the estimated entropy from many programmable switches, paving the way towards the estimation of network traffic entropy on a \emph{network-wide} scale \cite{ding2019incremental} to improve the statistical relevance of monitored values.

\vspace{-0.5em}
 \section{Conclusion and future work} \label{sec:conclusion}
In this paper, relying on recently-proposed logarithmic and exponential function estimation solutions, we presented P4LogLog to estimate the number of distinct flows in the network by only using P4-supported operations. We then proposed P4NEntropy, a strategy that leverages P4LogLog for the estimation of normalized network traffic entropy directly in the switch's data plane. Finally, P4DDoS has been designed on top of P4NEntropy, with the goal of detecting DDoS attacks by means of an entropy-based system.


We also evaluated all of our proposed approaches and compared them with state-of-the-art solutions. 
Results show that P4LogLog  has better accuracy than the state of the art especially when memory availability is small (i.e., smaller than 640 Bytes). 
Furthermore, P4NEntropy shows comparable accuracy on entropy estimation to existing approaches, but it leverages time-based observation windows (instead of fixed packet-based) and avoids the usage of TCAM (relying only on P4-supported operations). Finally, P4DDoS outperforms existing DDoS detection solutions implemented in P4 in terms of detection accuracy, especially in the case of internal Botnet DDoS attacks, while implementing a simpler logic. Moreover, unlike existing approaches in literature, all of our strategies avoid any communication overhead between controller and programmable switches, since they work entirely in the data plane. Specifically, P4DDoS only reports an alarm to the controller when an attack is detected.



As future work, we plan to extend our solution to detect other types of DDoS attacks, e.g. low-packet-rate, link flooding or carpet bombing attacks, with high accuracy. Furthermore, we also intend to work on an algorithm for the entropy-based detection of DDoS attacks on a network-wide scale, by collecting and combining the distributed entropy information from multiple programmable switches.

\vspace{-0.5em}
\end{document}